\documentclass[aps,prl,reprint,superscriptaddress,twocolumn,showkeys,amsmath,amssymb,longbibliography]{revtex4-1}
\usepackage[english]{babel}
\usepackage{amsmath,amssymb,bbm,mathrsfs,bm,braket,color,graphicx,comment,amsfonts,dsfont}
\usepackage[colorlinks,citecolor=blue,urlcolor=blue]{hyperref}
\usepackage[mathscr]{euscript}
\usepackage{color}

\newcommand{\pd}{{\phantom\dag}}

\begin{document}

\title{An anomalous higher-order topological insulator}

\author{S. Franca}

\affiliation{Institute for Theoretical Solid State Physics, IFW Dresden, 01171 Dresden, Germany}

\author{J. van den Brink}
\affiliation{Institute for Theoretical Solid State Physics, IFW Dresden, 01171 Dresden, Germany}
\affiliation{Institute for Theoretical Physics, TU Dresden,  01069 Dresden, Germany} 

\author{I. C. Fulga}
\affiliation{Institute for Theoretical Solid State Physics, IFW Dresden, 01171 Dresden, Germany}

\date{\today}
\begin{abstract}

Topological multipole insulators are a class of higher order topological insulators (HOTI) in which robust fractional corner charges appear due to a quantized electric multipole moment of the bulk. This bulk-corner correspondence has been expressed in terms of a topological invariant computed using the eigenstates of the Wilson loop operator, a so called ``nested Wilson loop'' procedure. We show that, similar to the unitary Floquet operator describing periodically driven systems, the unitary Wilson loop operator can realize ``anomalous'' phases, that are topologically non-trivial despite having a trivial topological invariant. We introduce a concrete example of an anomalous HOTI, which has a quantized bulk quadrupole moment and fractional corner charges, but a vanishing nested Wilson loop index. A new invariant able to capture the topology of this phase is then constructed. Our work shows that anomalous topological phases, previously thought to be unique to periodically driven systems, can occur and be used to understand purely time-independent HOTIs.

\end{abstract}

\maketitle

\textit{Introduction.}---In topological insulators, bulk-boundary correspondence relates the presence of robust boundary phenomena to a quantity determined from the bulk system, a topological invariant ~\cite{Thouless1982, Kane2005, Hasan2010, Qi2011, BernevigBook}. In strong topological insulators, the D-dimensional bulk is gapped and the topological invariant counts the number of gapless modes present on the $({\rm D}-1)$-dimensional boundaries of the crystal.
In the recently introduced higher-order topological insulators (HOTI) ~\cite{ Benalcazar2017, Benalcazar2017a, Langbehn2017, Hayashi2018, Song2017, Schindler2018, Schindler2018b, Wang2018, Ezawa2018, Ezawa2018a, Ezawa2018d,  Khalaf2018, Dwivedi2018a, Miert2018, Ezawa2018b, Hsu2018, Yan2018, Wang2018a, Trifunovic2018, Geier2018, Liu2018, Serra-Garcia2018, Serra-Garcia2018a, Peterson2018, Zhang2018, Imhof2018}
 however, both the bulk and the boundaries are gapped, so that standard bulk-boundary correspondence no longer applies. In some cases, the HOTI invariant determines the presence of topologically protected gapless modes on some regions of the boundary which have dimensions $({\rm D}-2)$ or less, such as the corners or hinges of a crystal  ~\cite{Song2017, Hayashi2018, Schindler2018, Schindler2018b, Wang2018, Ezawa2018, Ezawa2018a, Ezawa2018d, Khalaf2018, Dwivedi2018a, Miert2018, Sessi2016}.

Gapless modes on corners or hinges are not the only manifestations of topology in a HOTI, however.
Extending the notion of bulk-boundary correspondence to that of bulk-hinge or bulk-corner correspondence allows for a greater variety of boundary phenomena to be associated to a topologically non-trivial bulk. In seminal works, Benalcazar \emph{et al.} have introduced a class of HOTIs dubbed ``quantized electric multipole insulators'' ~\cite{Benalcazar2017, Benalcazar2017a}, whose non-trivial nature leads to topologically protected corner \emph{charges}, not states.
Among others, they considered a 2D topological quadrupole insulator (TQI), whose topological invariant is a bulk quadrupole moment, $q_{xy}$. The latter is quantized to $0$ (trivial) or $\frac{e}{2}$ (non-trivial) by lattice symmetries, with $e$ the electron charge. This leads to quantized tangential edge polarizations and fractional corner charges.
The defining relation of this HOTI is:
\begin{equation}\label{eq:top_quad_def}
 q_{xy}  = | p_x^{\rm edge} | =| p_y^{\rm edge} | = | Q_{\rm corner}|,
\end{equation}
where $p_x^{\rm edge}$ and $p_y^{\rm edge}$ are the tangential polarizations per unit length of the x and y edge and $Q_{\rm corner}$ is the fractional corner charge.
The above relations illustrate the bulk nature of the TQI, distinguishing it from phases in which the corner charge arises only due to ``free'' electric dipoles at the boundaries \cite{Benalcazar2017}. In the later case, $Q_{\rm corner} =  p_x^{\rm edge} + p_y^{\rm edge}$, violating Eq.~\eqref{eq:top_quad_def}.

To compute the bulk quadrupole moment, Refs.~\cite{Benalcazar2017,Benalcazar2017a} use the Wilson loop, a unitary operator whose spectrum represents the Wannier centers of electronic wavefunctions in the crystal. In the TQI these centers form gapped Wannier bands, and the quadrupole moment is determined from a topological invariant associated to these bands. An essential observation in the context of this work is that this novel procedure, termed a ``nested Wilson loop'' formalism, is unlike those used for conventional topological phases.
It determines the invariant from the eigenstates of a unitary operator and not directly from those of the Hermitian, Hamiltonian operator.

The topological phases of \emph{unitary} operators are well studied in the context of periodically driven systems ~\cite{Kitagawa2010, Lindner2011, Kitagawa2012, Nathan2015, Potter2016, Else2016, Rudner2013, Titum2016, Nathan2017, Kundu2017, Maczewsky2017, Mukherjee2017}, which are usually described in terms of the time evolution operator over one driving period -- the Floquet operator. Its unitary nature means the spectrum is $2\pi$ periodic, enabling so called ``anomalous topological phases'' ~\cite{Kitagawa2010, Rudner2013, Titum2016, Nathan2017, Kundu2017, Maczewsky2017, Mukherjee2017}. In the latter, the invariant associated to the Floquet operator vanishes, failing to capture the topologically non-trivial nature of the bulk. 
Anomalous topological phases so far have been hallmarks of periodically driven systems, and are considered impossible to achieve in a time-independent setting. Ultimately, however, their presence is possible solely due to the fact that non-trivial topology is realized using a unitary operator instead of a Hermitian one. Is it then possible for the unitary Wilson loop of a HOTI to host such anomalous phases?

In this work, we show that time-independent HOTIs can indeed host anomalous phases, introducing a class of systems we dub \emph{anomalous HOTIs}. Using the familiar language of Majorana bound states, we build a model of an electric insulator with a quantized bulk quadrupole moment and fractional corner charges. The system obeys the TQI relation Eq.~\eqref{eq:top_quad_def}, but has a vanishing nested Wilson loop invariant. We then adapt this formalism and formulate a new invariant which correctly captures the topologically non-trivial nature of the phase.

\begin{figure}[tb]
 \includegraphics[width=\columnwidth]{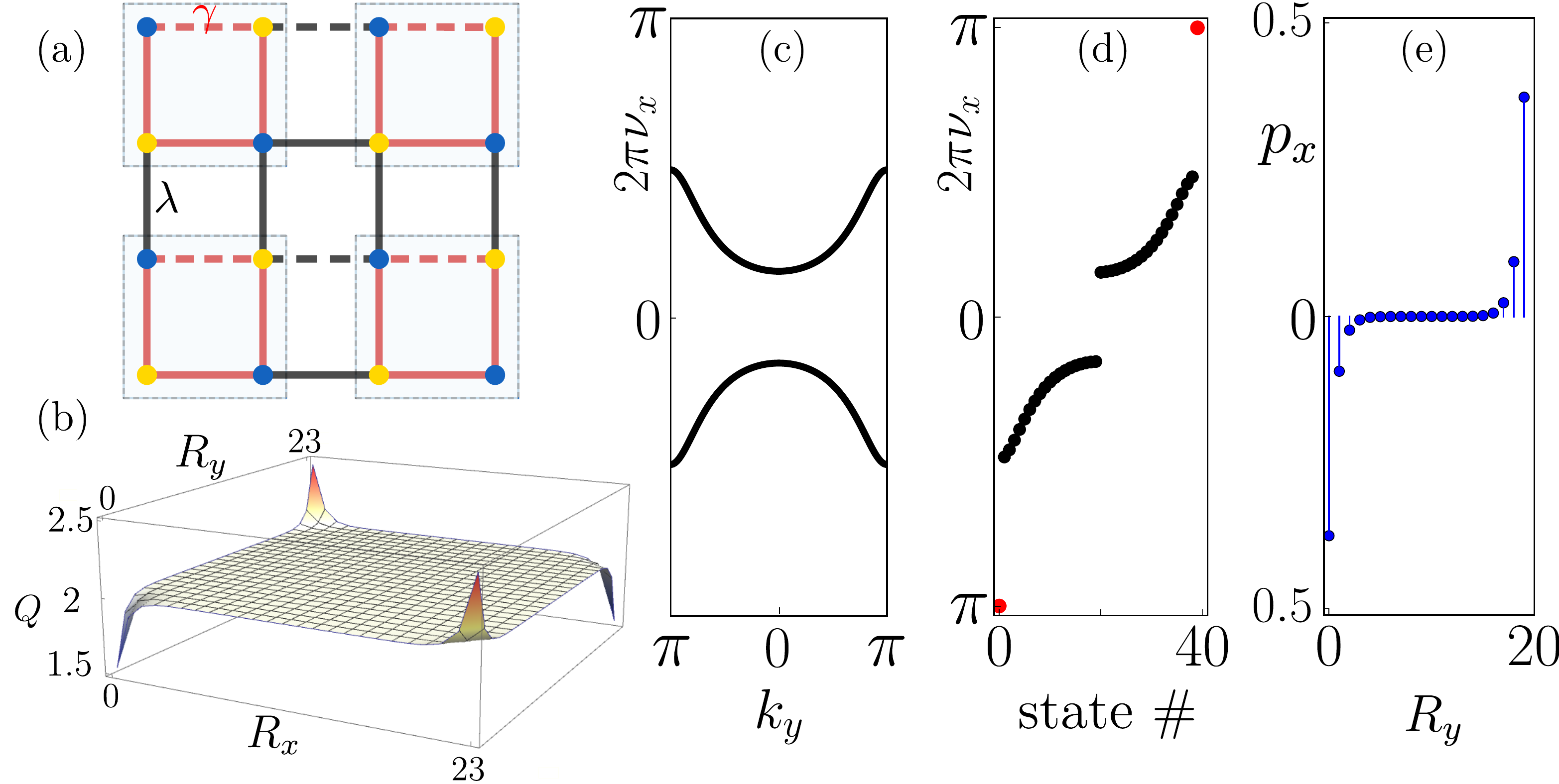}
  \caption{(a) TQI described by Eq.~\eqref{eq:BBHHam0}. Solid/dashed lines represent hoppings with positive/negative amplitudes, the red lines represent intracell hoppings $\gamma$, and the black ones intercell hoppings $\lambda$. Unit cells are marked by gray squares, with blue and yellow circles denoting sites belonging to different sublattices. (b) Charge density ($Q$) plot for a finite system consisting of $24\times24$ unit cells, where $R_x$ and $R_y$ label the unit cells in real space. The excess corner charge is $\pm1/2$. (c) Bulk Wannier bands of $\mathcal{W}_{x,{\bf k} = (\pi,\pi)}$, obtained by discretizing the BZ using $51$ k-points. (d) Wannier spectrum in the strip geometry (infinite in x, 20 unit cells in y). The $\nu=\pm0.5$ edge modes are shown in red. (e) Tangential polarization along $x$ 
as a function of position in the $y$-direction, $R_y$. The integrated polarization over half of the lattice sites yields $\pm 1/2$. We use $\gamma=0.5$ and $\lambda=1$. As explained in Refs.~\cite{Benalcazar2017, Benalcazar2017a}, to fix a sign for the polarization and the corner charges, we add a term $\delta \tau_z \sigma_0$ to Eq.~\eqref{eq:BBHHam0} ($\delta=10^{-3}$), which weakly breaks the two mirror symmetries ${\cal M}_x$ and ${\cal M}_y$ but not their product, the inversion symmetry ${\cal I} = {\cal M}_x{\cal M}_y$.
}
\label{fig:SSH}
\end{figure}

\textit{Nested Wilson loops.}---
We begin by briefly reviewing the previously introduced TQI and the nested Wilson loop procedure ~\cite{Benalcazar2017,Benalcazar2017a}.
We consider a system of spinless, non-interacting fermions on a square lattice with dimerized nearest-neighbor hoppings and a $\pi$ flux threading every plaquette (see Fig.~\ref{fig:SSH}a). Setting $\hbar=1$ and the lattice constant $a=1$, the Hamiltonian is
\begin{align}\label{eq:BBHHam0}
\begin{split}
h(\mathbf{k}) {} &= (\gamma  + \lambda \cos{k_x}) \tau_x \sigma_0 - \lambda \sin{k_x} \tau_y \sigma_z \\
& -  (\gamma  + \lambda \cos{k_y}) \tau_y \sigma_y - \lambda \sin{k_y} \tau_y \sigma_x,
\end{split}
\end{align}
where ${\bf k} = (k_x,k_y)$, $\tau$ Pauli matrices act on the sublattice degree of freedom, and $\sigma$'s parameterize the degree of freedom associated to the two sites within a sublattice. Intracell and intercell hoppings are $\gamma$ and $\lambda$, respectively.

For $|\gamma|<|\lambda|$, the system obeys all the requirements of a TQI, Eq.~\eqref{eq:top_quad_def}. The 2D bulk is gapped and each edge forms a non-trivial Su-Schrieffer-Heeger (SSH) chain \cite{Su1979}, leading to a single protected zero mode at every corner. At half filling, there are $N_\text{occ}=2$ occupied bands in the bulk and the four degenerate corner states contain two electrons in total, leading to a fractional corner charge $|Q_{\rm corner}|=1/2$ (see Fig.~\ref{fig:SSH}b), where we set the electron charge $e=1$ throughout the following.

To determine the polarizations, we define Wilson loop operators describing parallel transport of eigenstates along a closed path $p$ in the Brillouin zone (BZ). In the thermodynamic limit, the Wilson loop  ~\cite{Wilczek1984, Berry1984} is a path-ordered exponential (denoted $\overline\exp$)
\begin{equation}\label{eq:wloop}
\mathcal{W}_{p} = \overline\exp \left( -i \oint_{p} dp\cdot{\cal A}_{\bf k}  \right),
\end{equation}
where ${\cal A}_{\bf k}$ is the Berry connection of the occupied Hamiltonian eigenstates $|u_{\bf k}\rangle$. As such, ${\cal A}_{\bf k}$ is an $N_\text{occ}\times N_\text{occ}$ matrix with elements $[{\cal A}_{\bf k}]^{m n } =- i \langle u^m_{\bf k} | \nabla^\pd_{\bf k} u^n_{\bf k} \rangle$. We only consider Wilson loops computed on non-contractible paths of the BZ (so called large Wilson loops). We define ${\cal W}_{x,{\bf k}}$ on a path along $k_x$, from ${\bf k}$ to ${\bf k}+(2\pi,0)$, and similarly ${\cal W}_{y, {\bf k}}$ from ${\bf k}$ to ${\bf k} + (0,2\pi)$, where ${\bf k}$ is called the base point of the Wilson loop. Diagonalizing these unitary operators,
\begin{equation}\label{eq:wloop_diag}
 \mathcal{W}_{x,\mathbf{k}} \ket{\nu_{x,\mathbf{k}}^j} = \exp\left[{i 2\pi \nu_x^j(k_y)}\right]  \ket{\nu_{x,\mathbf{k}}^j},
\end{equation}
where $j\in\{1,\ldots,N_\text{occ}\}$, yields eigenstates $\ket{\nu_{x,\mathbf{k}}^j}$ with components $[\nu_{x,\mathbf{k}}^j]^n$ as well as eigenphases $\nu_x^j$. Note that while in general eigenstates depend explicitly on the base point ${\bf k}$, the eigenvalues are independent of the momentum along the path. As such, $\nu_x^j$ is only a function of $k_y$, while the eigenphase of ${\cal W}_{y, {\bf k}}$, $\nu_y^j$, only depends on $k_x$. Since the model Eq.~\eqref{eq:BBHHam0} has a four-fold rotation symmetry, we will only focus on ${\cal W}_{x, {\bf k}}$ in the remainder of this section, with the understanding that ${\cal W}_{y, {\bf k}}$ gives identical results.

We numerically determine the Wilson loop operators by discretizing the BZ. The procedure is thoroughly explained in Refs. ~\cite{Benalcazar2017,Benalcazar2017a}, so we do not repeat it here. In the Supplemental Material however \cite{SuppMat2}, we give a detailed account of these steps and include the code library we have developed for this purpose, which is general enough to be used for a large variety of HOTIs. 
The two eigenphases $\nu_x^j$ obtained from Eq.~\eqref{eq:wloop_diag} are shown in Fig.~\ref{fig:SSH}c. They form gapped Wannier bands which are positioned symmetrically around 0, which is a consequence of the two mirror symmetries, $\mathcal{M}_x = \tau_x \sigma_z$ and $\mathcal{M}_y = \tau_x \sigma_x$, acting in the x and y directions, respectively. Since the bulk polarization is given by the sum of Wilson loop eigenphases, $p_x (k_y) = \sum_j \nu_x^j(k_y)$, it vanishes at every momentum, a necessary requirement for a TQI ~\cite{Benalcazar2017,Benalcazar2017a}.
Of course, due to four-fold rotation the same result holds for $p_y(k_x)$.

The key insight behind the nested Wilson loop formalism is to notice that, since the Wannier bands are gapped, they carry their own topological invariants. Non-trivial Wannier bands then lead to topologically protected Wannier modes at the boundaries of the system. 
To introduce boundaries, we consider Eq.~\eqref{eq:BBHHam0} in a strip geometry, infinite along $k_x$ and containing 20 unit cells in the y direction. From the strip Hamiltonian $h(k_x,R_y)$, where $R_y$ labels the unit cells, we compute the large Wilson loop along the only remaining momentum, $k_x$, and show its eigenphases in Fig.~\ref{fig:SSH}d. The gapped Wannier centers are accompanied by topological modes (shown in red) which are pinned to $\nu_x=\pm0.5$ by mirror symmetry. These are localized on opposite boundaries of the system, leading to a quantized edge polarization. We confirm this in Fig.~\ref{fig:SSH}e, which shows the tangential polarization $p_x$ in the strip geometry. While there is no polarization in the bulk, the boundaries have a quantized polarization $\pm 1/2$, which accompanies the fractional corner charges.

Since $| p_x^{\rm edge} | =| p_y^{\rm edge} | = | Q_{\rm corner}|=1/2$, the origin of the corner charges must be due to a bulk quadrupole moment, $q_{xy}=1/2$. To compute the latter, Refs.~\cite{Benalcazar2017,Benalcazar2017a} use the topological invariants associated to the Wannier bands in Fig.~\ref{fig:SSH}c. They split the Wannier bands into two sectors, an ``occupied'' and an ``unoccupied'' one:
\begin{equation}\label{eq:occ_unocc}
 \begin{split}
  \nu_x^- &= \{ \nu_x^j(k_y) \,\text{ such that }\, \nu_x^j(k_y) < 0 \}, \\
  \nu_x^+ &= \{ \nu_x^j(k_y) \,\text{ such that }\, \nu_x^j(k_y) > 0 \}.
 \end{split}
\end{equation}
Notice that, similar to Floquet systems, this distinction is not well defined, since the whole spectrum is $2\pi$ periodic, and there is no notion of a band being ``above'' or ``below'' another. Nevertheless, this splitting allows to separate the space of occupied Hamiltonian eigenstates into two Wannier band subspaces, corresponding to Wannier states
\begin{equation}\label{eq:wannier_subspace}
\ket{w_{x,\mathbf{k}}^{\pm,r}}  = \sum_{n=1}^{N_\text{occ}} \ket{u_{\mathbf{k}}^n} [\nu_{x,\mathbf{k}}^{\pm,r}]^n
\end{equation}
Here, the superscript $\pm$ denotes the Wannier band subspace, with $r\in\{1,\ldots,N_\text{occ}/2\}$ labeling the bands within a subspace. The non-trivial nature of each of the Wannier subspaces is then determined from the Wannier sector polarization. In the thermodynamic limit,
\begin{equation}\label{eq:FinalPol}
p_y^{\nu_x^\pm} = - \frac{1}{(2\pi)^2} \int_{BZ} {\rm Tr}\left[ \tilde{\mathcal{A}}_{y,\mathbf{k}}^{\nu_x^\pm} \right] d^2\mathbf{k}
\end{equation}
is a $\mathbb{Z}_2$ topological index, $p_y^{\nu_x^\pm}\in\{0,1/2\}$,
where $[\tilde{\mathcal{A}}_{y,{\bf k}}^{\nu_x^\pm}]^{mn} = - i \bra{w_{x,{\bf k}}^{\pm,m}} \partial_{k_y} \ket{w_{x,{\bf k}}^{\pm,n}}$ are the matrix elements of the $(N_\text{occ}/2)\times (N_{\rm occ}/2)$ Berry connection of the $\pm$ Wannier subspace. In the case of Eq.~\eqref{eq:BBHHam0}, $N_\text{occ}=2$ so the connection is a scalar. We find $p_y^{\nu_x^\pm} =p_x^{\nu_y^\pm}=1/2$ in agreement with Refs.~\cite{Benalcazar2017,Benalcazar2017a}, which define the bulk quadrupole invariant as,
\begin{equation}\label{eq:qxy}
q_{xy} =p_y^{\nu_x^{+}} p_x^{\nu_y^{+}} +p_y^{\nu_x^{-}} p_x^{\nu_y^{-}},
\end{equation}
obtaining a quantized value $q_{xy}=1/2$.

\emph{Anomalous HOTI.}---
In a HOTI, non-trivial Wannier sector invariants, Eq.~\eqref{eq:FinalPol}, imply topological Wannier edge modes, leading to quantized edge polarizations and fractional corner charges. The converse statement is however not true. As we show in the following, due to the unitary nature of the Wilson loop, Wannier edge modes can occur even if the subspaces $\nu_x^{\pm}$ have trivial invariants.

\begin{figure}[tb]
\includegraphics[width=\columnwidth]{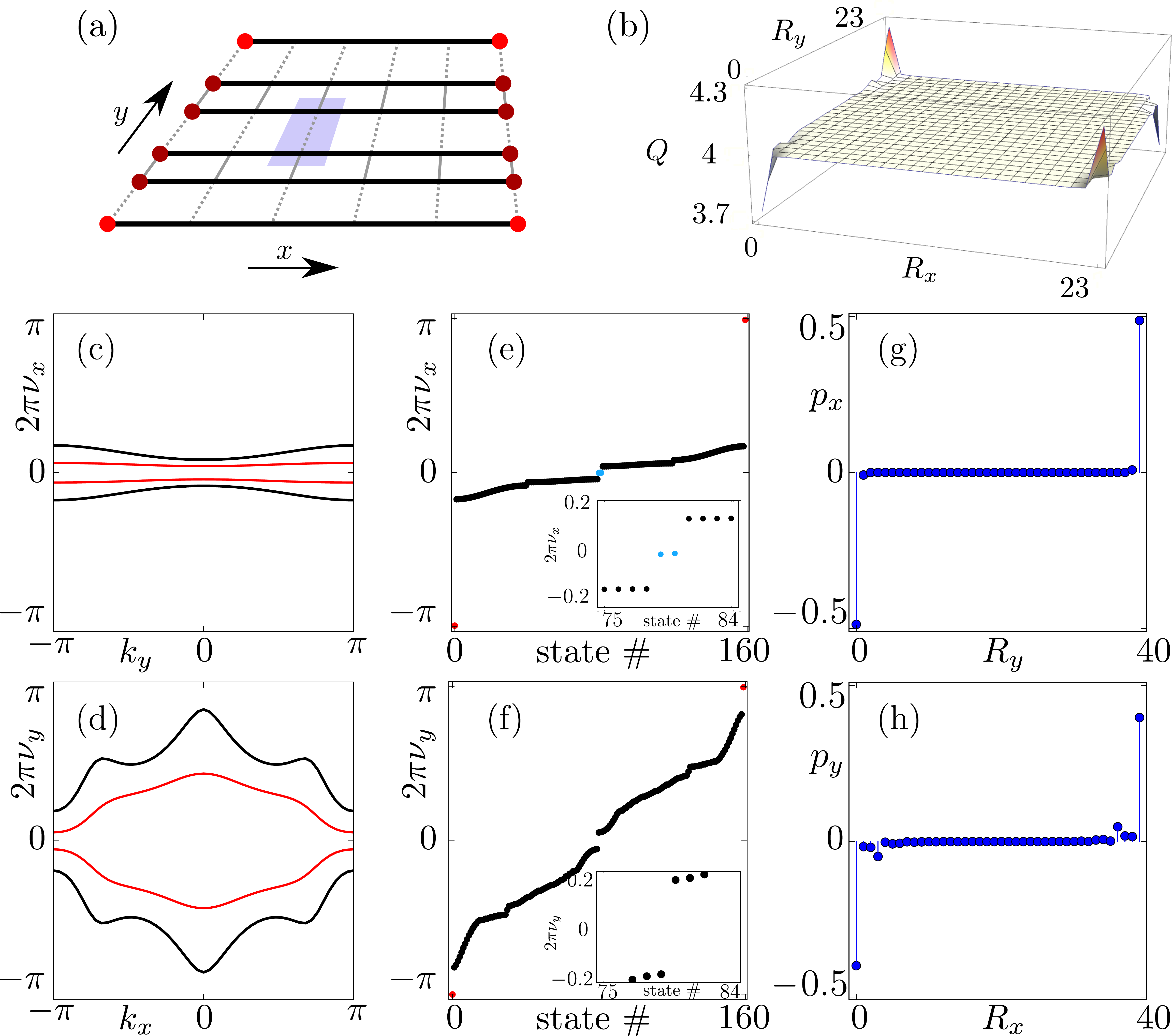}
\caption{(a) Array of Majorana wires (black lines), coupled in a dimerized fashion (gray solid/dashed lines). The unit cell (blue) contains two wires. Majorana modes which are gapped out by the dimerized coupling are shown in dark red, whereas protected corner modes are shown in bright red. (b) Charge density plot of a system of $24 \times 24$ unit cells. The integrated excess charge is $\pm1/2$ for each corner. (c, d) Bulk Wannier bands of $\mathcal{W}_{x,(\pi,\pi)}$ and $\mathcal{W}_{y,(\pi,\pi)}$. (e, f) Wannier spectra in a strip geometry, infinite along either $k_x$ (e) or $k_y$ (f), with 40 unit cells  in the finite direction. Both spectra show edge modes at $\nu=\pm0.5$ (red), but the strip along $k_x$ also shows topological 0-modes (blue). The insets show closeups of the gaps around $\nu=0$.
(g, h) The corresponding edge polarization is quantized to $\pm1/2$ in both cases. To fix the sign of the charge and polarization, we add a mirror symmetry breaking term $\delta\tau_y \sigma_y \eta_z$ ($\delta=10^{-2}$).}  
\label{fig:MHam}
\end{figure}

To model an anomalous HOTI, we consider a 2D array of Majorana nanowires \cite{Oreg2010, Lutchyn2010}. The wires are coupled to each other in a dimerized fashion, such that their end modes gap out in pairs, leaving topologically protected zero modes only at the corners (see Fig.~\ref{fig:MHam}a). In this respect, the system is very similar to the TQI of Eq.~\eqref{eq:BBHHam0}: each corner state is a simultaneous topological mode of two non-trivial edges. The horizontal edges of Fig.~\ref{fig:MHam}a are topological nanowires, while the vertical edges are Kitaev chains in the non-trivial phase \cite{Kitaev2001}.
The Hamiltonian reads
\begin{align} \label{eq:MFHam}
\begin{split}
& H(\mathbf{k}) {} = [2t_x(1- \cos{k_x})  -\mu]\tau_z \sigma_0 \eta_0  + \\ 
&   V_z \tau_0 \sigma_z \eta_0 +  \Delta \tau_x \sigma_0 \eta_0 + \alpha \sin{k_x} \tau_z \sigma_y \eta_0  -  \\
& \beta_1 \tau_z \sigma_x \eta_y - \beta_2 \sin{k_y} \tau_z \sigma_x \eta_x + \beta_2 \cos{k_y} \tau_z \sigma_x \eta_y
\end{split}
\end{align}
where $t_x$ and $\alpha$ are the nearest-neighbor hopping strength and the spin-orbit coupling (SOC) strength in the x direction (along the wires), $\mu$ is the chemical potential, $V_z$ is the Zeeman energy, and $\Delta$ is the superconducting pairing strength. In the y direction, $\beta_{1}<\beta_{2}$ are dimerized Rashba SOC terms connecting neighboring wires within and between unit cells, whereas Pauli matrices 
$\tau$, $\sigma$, and $\eta$ act on the particle-hole, spin, and wire space, respectively.
We set $\alpha=3.7$, $t_x=1.7$, $\mu=-0.9$, $\Delta=1.6$, $V_z=2.7$,
$\beta_{1}=0.8$, and $\beta_{2}=6.2$ throughout the following.

As is conventional when describing the topology of superconductors \cite{Teo2010, Budich2013}, in the following we neglect the BdG nature of the Hamiltonian Eq.~\eqref{eq:MFHam}, treating it instead as a charge-conserving Bloch Hamiltonian with a well-defined filling. As such, from now on we consider each of the four corner modes \emph{not} as a Majorana bound state, but an an electron state which may be filled independently of the others. At half filling, there are $N_\text{occ}=4$ occupied bulk bands, and each corner mode is half filled, such that $|Q_{\rm corner}|=1/2$ (see Fig.~\ref{fig:MHam}b), the first indication of a TQI.

We follow the procedure summarized in the previous section, and determine the bulk and edge polarizations. Since Eq.~\eqref{eq:MFHam} lacks rotation symmetry, we show the eigenphases of both ${\cal W}_{x,(\pi,\pi)}$ and ${\cal W}_{y,(\pi,\pi)}$ in Fig.~\ref{fig:MHam}c, d. There are four Wannier bands in total, such that each Wannier sector [Eq.~\eqref{eq:occ_unocc}] contains two bands, shown in red and black. Similarly to the TQI of Eq.~\eqref{eq:BBHHam0}, the bands are gapped and positioned symmetrically around 0 due to mirror symmetries (here, $\mathcal{M}_x = \eta_z \sigma_z $ and $\mathcal{M}_y = \eta_y$), which leads to a vanishing bulk polarization. The behavior of ${\cal W}_{y,(\pi,\pi)}$ is identical to that of the previous model: topological Wannier edge modes appear at $\nu_y=\pm0.5$ (Fig.~\ref{fig:MHam}f), leading to edge polarizations which are quantized to $\pm1/2$ (Fig.~\ref{fig:MHam}h).

The crucial difference with respect to Eq.~\eqref{eq:BBHHam0} is given by the Wannier spectrum associated to ${\cal W}_x$ (Fig.~\ref{fig:MHam}e). In a strip geometry infinite along $k_x$, the Wilson loop shows two \emph{different} kinds of edge modes. The first (shown in red) corresponds to $\pi$-eigenphases of the Wilson loop, $\nu=\pm0.5$, leading to quantized edge polarizations. The second kind of topological edge mode (shown in blue) has $\nu=0$ instead. Both 0- and $\pi$-modes are compatible with the mirror symmetry, which renders the spectrum $\pm\nu$ symmetric. However, since the polarization is given by the eigenphases of the Wilson loop, the zero modes do not contribute to the edge polarizations, even though their eigenstates are localized on the boundaries of the system \cite{SuppMat2}.

Fig.~\ref{fig:MHam}e is analogous to the spectrum of a 1D Floquet topological phase, where it is known that two kinds of protected modes can occur, at 0 and at $\pi$ quasi-energies, respectively \cite{Jiang2011, Kundu2013, Tong2013, Yao2017, Yang2018}. When both types of topological boundary states are present in the same system, the resulting phase is termed ``anomalous'', since the topological invariant associated to the bulk Floquet operator vanishes. The same occurs for the Wilson loop operators of Fig.~\ref{fig:MHam}. For ${\cal W}_{y,(\pi,\pi)}$, only $\pi$-modes are present so the Wannier sector polarizations [Eq.~\eqref{eq:FinalPol}] are $p_x^{\nu_y^\pm}=1/2$ as expected. For ${\cal W}_{x,(\pi,\pi)}$ on the other hand, we find trivial Wannier subspace invariants, $p_y^{\nu_x^\pm}=0$.

Taken by themselves, the Wannier sector invariants would indicate a trivial HOTI, $q_{xy}=0$, as per Eq.~\eqref{eq:qxy}. We know however that this is not the case, since both of the bulk Wilson loops show topological edge modes when boundaries are introduced. These modes lead to quantized edge polarizations equal to the corner charge, $| p_x^{\rm edge} | =| p_y^{\rm edge} | = | Q_{\rm corner}|=1/2$, which is the defining relation of a TQI, Eq.~\eqref{eq:top_quad_def}. To overcome this discrepancy, we adapt the nested Wilson procedure and introduce a new bulk index. The key observation is that for the model Eq.~\eqref{eq:MFHam} there are two Wannier bands in each of the $\nu_x^\pm$ subspaces, such that the associated Berry connection of each sector, ${\cal \tilde{A}}_{y, {\bf k}}^{\nu_x^\pm}$, is a $2\times2$ matrix. This means that the Wannier sector polarization Eq.~\eqref{eq:FinalPol} effectively sums the invariants of the two Wannier bands, so that two non-trivial bands lead to a vanishing topological index. We take this into account and compute the topological index of each band \emph{separately}, splitting the trace in Eq.~\eqref{eq:FinalPol} into two separate integrals, $p_y^{\nu_x^{\pm}}=p_y^{\nu_x^{\pm}, 1}+p_y^{\nu_x^{\pm}, 2}$, where the superscript $1,2$ denotes the index of the black and red Wannier bands of Fig.~\ref{fig:MHam}c, d. Note that since the Wannier bands do not cross, the index of each band is well defined and quantized to 0 or $1/2$ \textcolor{red}{\cite{SuppMat2}}. This allows us to redefine the bulk quadrupole index of Eq.~\eqref{eq:qxy} as
\begin{equation}\label{eq:InvNew}
q_{xy} = \sum_{r=1}^{N_\text{occ}/2} p_y^{\nu_x^{+},r} p_x^{\nu_y^{+},r} + p_y^{\nu_x^{-},r} p_x^{\nu_y^{-},r}.
\end{equation}
We find that in each Wannier sector both Wannier bands of ${\cal W}_{x,{\bf k}}$ are non-trivial $p_y^{\nu_x^{\pm},1} =p_y^{\nu_x^{\pm},2}=1/2$. 
On the other hand, for ${\cal W}_{y,{\bf k}}$ only $p_x^{\nu_y^{\pm},1}$ is non-zero, such that Eq.~\eqref{eq:InvNew} gives $q_{xy}=1/2$ signaling a non-trivial TQI. 
In the Supplemental Material we further confirm these values of the topological invariants by showing the phase transitions that occur between Wannier bands as a function of model parameters $\beta_{1,2}$, $V_z$ and $\mu$.
Note that since $| p_x^{\rm edge} | =| p_y^{\rm edge} | = | Q_{\rm corner}|=1/2$, as shown in Fig.~\ref{fig:MHam}, the index $q_{xy}=1/2$ defined in Eq.~\eqref{eq:InvNew} represents the physical quadrupole moment of this system.

\emph{Conclusion.}---
We have shown that anomalous topological phases, previously considered unique to periodically-driven systems, can occur in time-independent HOTIs. We have introduced an example of such an anomalous HOTI, in which the unitary Wilson loop has topological properties analogous to those of an anomalous Floquet operator. Topological boundary states appear both at 0 and at $\pi$ values of the eigenphase, leading to a system with quantized edge polarizations and corner charges, but in which the nested Wilson loop index [Eqs.~\eqref{eq:FinalPol} and \eqref{eq:qxy}] vanishes. A new bulk invariant has been introduced [Eq.~\eqref{eq:InvNew}], which takes into account the topological properties of each individual Wannier band, as opposed to those of the entire Wannier sector.

There are however important differences between the topology of a Wilson loop and that of the Floquet operator describing a driven system. In Floquet systems, observing topological phases is often hindered by the requirement of filling specific bands \cite{DAlessio2015}. Further, when interactions are present, Floquet band populations are known to evolve towards a featureless infinite temperature state \cite{DAlessio2014, Lazarides2014}, unless the system is many body localized. In contrast, in static HOTIs there is no notion of ``filling'' for the bands of the Wilson loop. The latter simply describe the positions of ground-state quasi-particles relative to the unit cells, and can be defined also in interacting systems \cite{Ortiz1994, Lee2008}. As such, HOTIs may provide a way to observe anomalous phases, while mitigating the above-mentioned difficulties.

Our work bridges the gap between the study of topological phases in static and time-periodic systems, and as such opens many new directions of future research. For instance, anomalous HOTIs can be extended to higher multipole moments, and we conjecture that a 3D array of coupled nanowires would realize a system with corner charges and an anomalous octupole index. Further, it is interesting to consider whether both 0- and $\pi$- modes can occur in a system with only two Wannier bands, like the one of Fig.~\ref{fig:SSH}. In that case, neither the index of Eq.~\eqref{eq:qxy} nor that of Eq.~\eqref{eq:InvNew} would be adequate to characterize the non-trivial nature of the phase, prompting the search for other topological invariants. Finally, we remark that anomalous phases are not restricted to insulating systems, and they should be possible also in higher-order topological semimetals. In fact, the gapless Wannier spectrum shown in Fig.~4d of Ref.~\cite{Lin2017} shows protected Dirac cones occurring simultaneously at $\nu=0$ and $\nu=\pm1/2$, a behavior usually associated to anomalous Floquet semimetals \cite{Zhou2016, Bomantara2016, Wang2016, Higashikawa2018}.

\begin{acknowledgments}

We thank Ulrike Nitzsche for technical assistance.

\end{acknowledgments}

\bibliography{AHOTI}

\end{document}

\title{Supplemental material}

\author{S. Franca}

\affiliation{Institute for Theoretical Solid State Physics, IFW Dresden, 01171 Dresden, Germany}

\author{J. van den Brink}
\affiliation{Institute for Theoretical Solid State Physics, IFW Dresden, 01171 Dresden, Germany}
\affiliation{Institute for Theoretical Physics, TU Dresden,  01069 Dresden, Germany} 

\author{I. C. Fulga}
\affiliation{Institute for Theoretical Solid State Physics, IFW Dresden, 01171 Dresden, Germany}

\date{\today}
\begin{abstract}
In this Supplemental Material, we provide more details on the models discussed in the main text as well on how to determine quantities of the defining relation for a TQI phase. In Appendix 1, we describe the 2D SSH model, including the phase transitions of Wannier bands. The second part of this Appendix is devoted to the model of coupled nanowires where we examine how gap closings of Wannier bands serve as a tool to predict their topological invariants. Furthermore, we also show the probability distribution of Wannier $0$- and $\pi$-modes and prove that both kinds of modes are edge localized. 
Appendix 2 is divided into three sections. In the first section, we present the calculation of the Wilson loop operator in the discrete limit in the bulk and discuss the basis fixing procedure. The calculation of the tangential edge polarization for the strip geometry is explained in the second section. The last part deals with a finite size system in which the charge density distribution reveals the presence of quantized corner charges.The quantization of the topological invariant is detailed in Appendix 3. At the end, Appendix 4 provides an analytical proof that the corner mode of the anomalous HOTI is the simultaneous end state of both edge Hamiltonians. 

\end{abstract}
% \pacs{...}

\maketitle

\section{Appendix 1: Models}

\subsection{Two-dimensional SSH model}
\begin{figure}[tb]
\begin{center}
\includegraphics[width=\columnwidth]{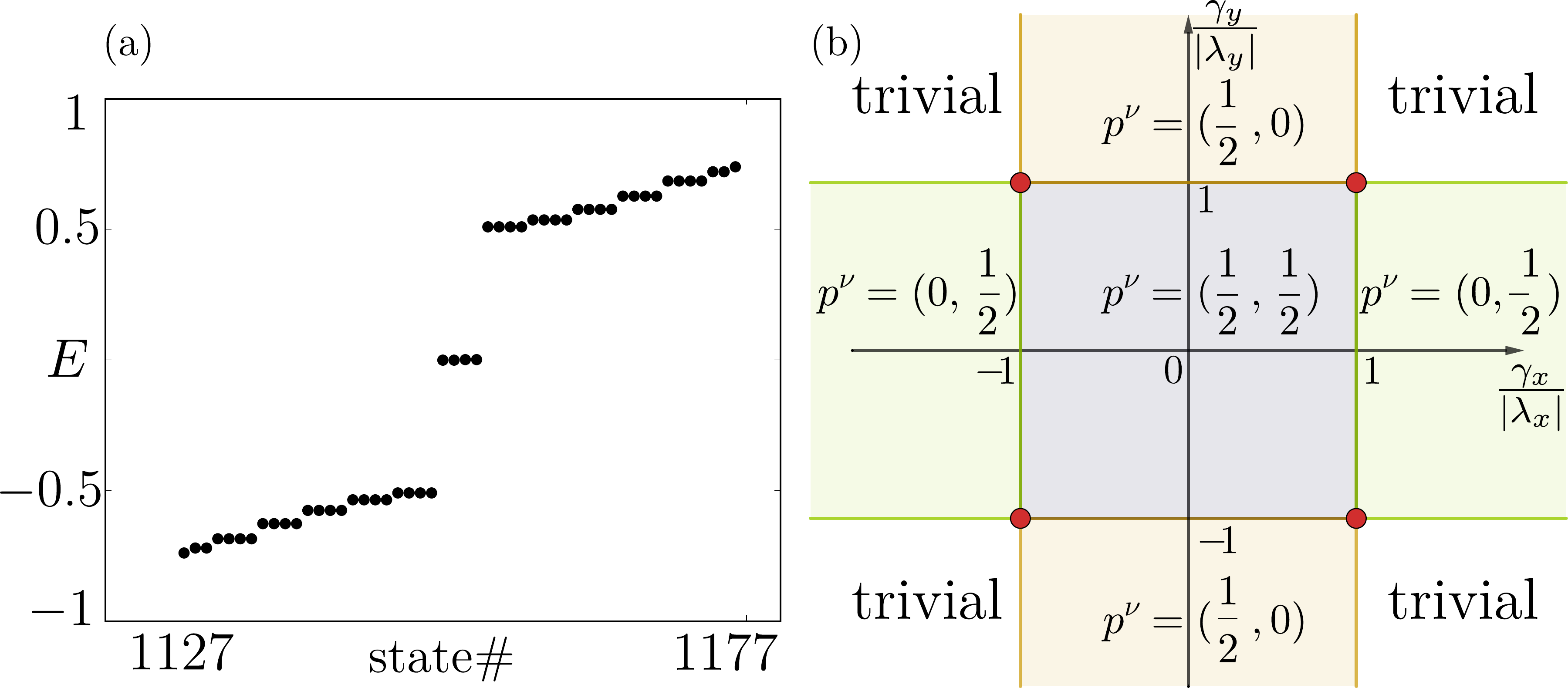}
\end{center}
\caption{ (a) The energy spectrum in the case of a TQI ($\gamma_x=\gamma_y=1/2$ and $\lambda_x=\lambda_y=1$), for a system of $24\times 24$ unit cells. (b) The phase diagram of a Hamiltonian given by Eq.~\eqref{eq:BBHHam}. The region of a TQI is colored in blue. Red points represent direct topological phase transitions from a HOTI to a trivial insulator. Such phase transitions are marked by the closing of the bulk gap. Regions colored in orange denote a region with a non-vanishing Wannier sector polarization in the $k_x$ direction while the regions colored in green have a non-vanishing Wannier sector polarization in the $k_y$ direction.  } 
\label{fig:PhaseDiagSSH}
\end{figure}

The Hamiltonian of a 2D SSH model is
\begin{align} \label{eq:BBHHam}
\begin{split}
H(\mathbf{k}) {} &= (\gamma_x  + \lambda_x \cos{k_x}) \tau_x \sigma_0 - \lambda_x \sin{k_x} \tau_y \sigma_z \\
& -  (\gamma_y  + \lambda_y \cos{k_y}) \tau_y \sigma_y - \lambda_y \sin{k_y} \tau_y \sigma_x,
\end{split}
\end{align}
where $\tau$'s are Pauli matrices representing sublattice degrees of freedom and $\sigma$'s are Pauli matrices that represent degrees of freedom associated to the two sites within the sublattice. Intracell hoppings are $\gamma_x$ and $\gamma_y$ while the intercell hoppings are $\lambda_x$ and $\lambda_y$. The eigenenergies are: 
\begin{equation*}
E^2 = \gamma_x^2+\lambda_x^2+2\gamma_x\lambda_x \cos{k_x}  +  \gamma_y^2+\lambda_y^2+2\gamma_y\lambda_y \cos{k_y}.
\end{equation*}
When at half filling, this model represents an insulator with two filled bands and with a well defined gap throughout the  Brillouin zone (BZ) except at the points of a topological phase transition directly to a trivial phase. We see that there are four possible points in the BZ where the gap closings can occur: at $(k_x,k_y)=(0,0)$ and $(\pi,\pi)$ when $\gamma_x=\mp\lambda_x$ and $\gamma_y=\mp\lambda_y$, or at $(k_x,k_y)=(0,\pi)$ and $(\pi,0)$ when $\gamma_x=\mp\lambda_x$ and $\gamma_y=\pm\lambda_y$. In the open geometry, the topological phase is characterized by four zero energy states separated from the bulk by a gap (Fig.~\ref{fig:PhaseDiagSSH}a).

\begin{figure}[tb]
\begin{center}
\includegraphics[width=\columnwidth]{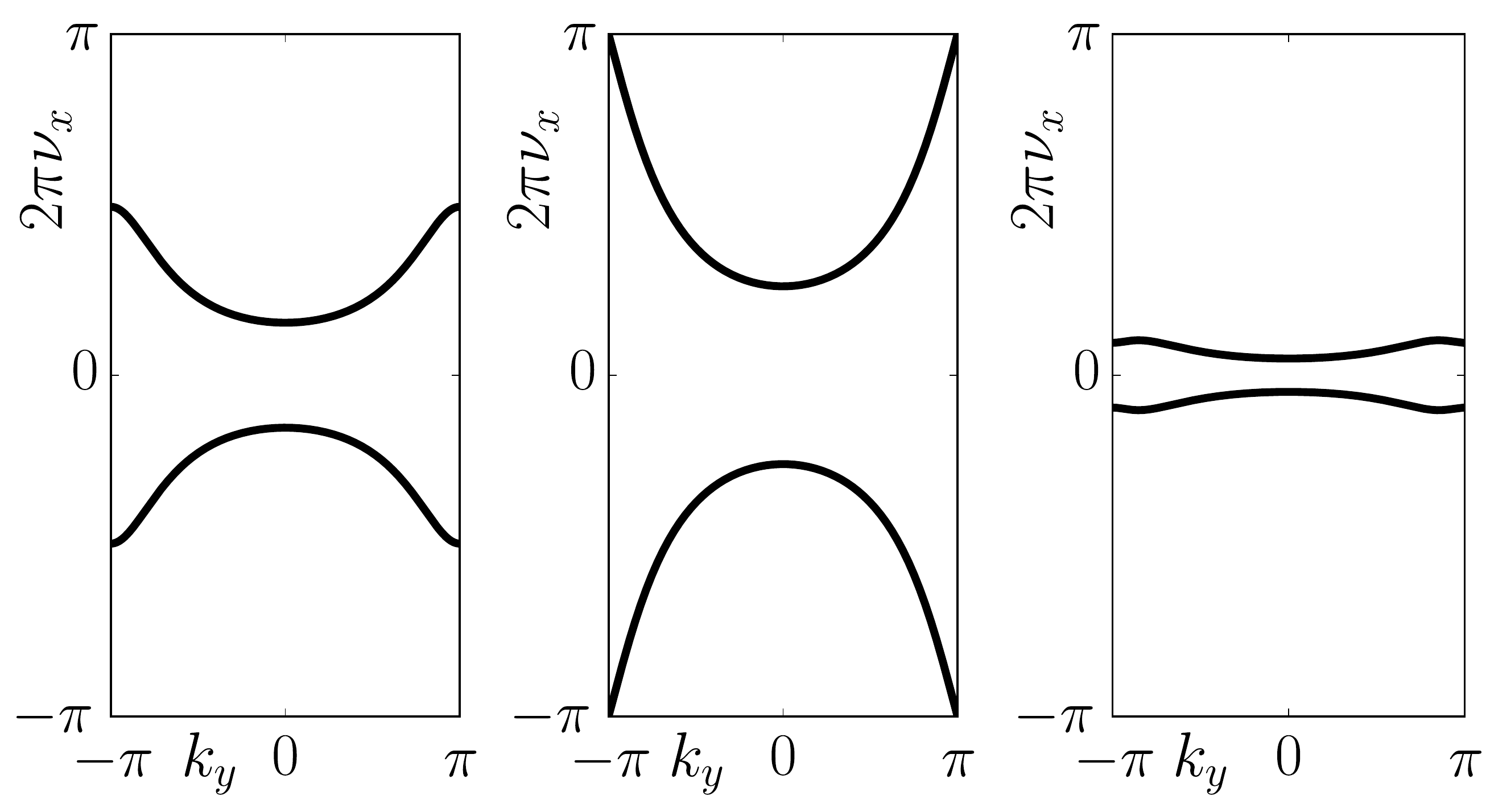}
\end{center}
\caption{The left panel represents Wannier bands in a HOTI phase described by the parameters $\gamma_x=\gamma_y=0.5$ and $\lambda_x = \lambda_y = 1$ while the central panel shows Wannier bands at the transition line between a HOTI and a phase $p^{\nu}=(1/2,0)$. The parameters are $\gamma_x=\gamma_y=0.5$, $\lambda_x=1$ and $\lambda_y=0.5$. Finally, the right panel represents a topologically trivial phase described by the parameters $\gamma_x=\gamma_y=1.5$ and $\lambda_x=\lambda_y=1$. The initial point of the Wilson loop is at $(k_x,k_y)=(\pi,\pi)$ and the BZ is discretized using $251$ k-points in each direction.} 
\label{fig:SSHTransition}
\end{figure}

The bulk gap closing points are represented on the phase diagram for a 2D SSH model in Fig.~\ref{fig:PhaseDiagSSH}b. Unlike the Hamiltonian given in the main text, Hamiltonian defined in Eq.~\eqref{eq:BBHHam} is not rotationally invariant since it allows for different hopping strengths in the x and y directions. This, in turn, results in a much richer phase diagram than for the case of a Hamiltonian with  a $\mathcal{C}_4$ symmetry, that is restored  once $\gamma_x =\pm \gamma_y$ and $\lambda_x =\pm \lambda_y$. As shown in Fig.~\ref{fig:PhaseDiagSSH}b, there are phases with only one non-vanishing Wannier sector polarization. These phases can be reached through phase transitions \cite{Benalcazar2017} in the Wannier spectrum.
For example, the Wannier phase transition depicted in Fig.~\ref{fig:SSHTransition} separates two phases in which $p^{\nu}= (1/2,1/2) \equiv q_{xy}=1/2$ and $p^{\nu}= (1/2,0)$, respectively. Here, $p^{\nu}$ stands for $p^{\nu^{\pm}}=(p_x^{\nu^{\pm}_y},p_y^{\nu^{\pm}_x})$ where $\pm$ labels the Wannier sector and $p_x^{\nu^{\pm}_y}$ denotes Wannier sectors' polarizations in the $k_x$ direction, obtained from diagonalizing Wilson loop operators calculated in the $k_y$ direction.

In this model, since there is only one Wannier band per sector, the Wannier sector polarization is equivalent to the polarization of the respective Wannier band. In the case of a system with several bands per Wannier sector, we replace the Wannier sector polarization $p_x^{\nu^{\pm}_y}$ by $(p_x^{\nu^{\pm}_{y},1},p_x^{\nu^{\pm}_{y},2},...)$ where $p_x^{\nu^{\pm}_{y},j}$ labels the polarization of a band $j$.

The gap closing in the Wannier spectrum shown in Fig.~\ref{fig:SSHTransition} results in a change of the bands' topological indexes. At the same time, the Wannier spectrum obtained from diagonalizing the other Wilson loop does not show any phase transitions for those parameter values. The consequence of a non-trivial Wannier sector in the bulk is the presence of edge modes in the Wannier spectrum of the strip Hamiltonian \cite{Benalcazar2017a}. At the transition to a phase with trivial Wannier bands, these edge modes are pushed into the bulk. 

\subsection{Anomalous HOTI}
The Hamiltonian of an anomalous HOTI reads
\begin{align} \label{eq:MFHam}
\begin{split}
& H(\mathbf{k}) {} = [2t_x(1- \cos{k_x})  -\mu]\tau_z \sigma_0 \eta_0  + \\ 
&   V_z \tau_0 \sigma_z \eta_0 +  \Delta \tau_x \sigma_0 \eta_0 + \alpha \sin{k_x} \tau_z \sigma_y \eta_0  -  \\
& \beta_1 \tau_z \sigma_x \eta_y - \beta_2 \sin{k_y} \tau_z \sigma_x \eta_x + \beta_2 \cos{k_y} \tau_z \sigma_x \eta_y,
\end{split}
\end{align}
where $t_x$ is the nearest-neighbor hopping strength in the x direction, $\mu$ is the chemical potential, $V_z$ is the Zeeman energy, $\Delta$ is the superconducting pairing strength,  $\alpha$ is the strength of Rashba SOC in the x direction and $\beta_1$/ $\beta_2$ are the strengths of Rashba SOC in the  y direction. We define two quantities used in the code: $\beta_{avg}$ and $\beta_{dim}$ such that $\beta_1= \beta_{avg} - \beta_{dim}$ and $\beta_2=\beta_{avg} + \beta_{dim}$. The matrices $\tau$, $\sigma$ and $\eta$ are Pauli matrices acting in the particle-hole, spin and wire spaces, respectively. 

The Hamiltonian has mirror symmetries $\mathcal{M}_x = \eta_z \sigma_z $ and $\mathcal{M}_y = \eta_y$ that anti-commute, inversion symmetry $\mathcal{I} = \eta_x \sigma_z $, particle-hole 
$\mathcal{P} = \tau_y \sigma_y \mathcal{K}$, chiral $\mathcal{C} = \tau_y \eta_z \sigma_y $, and time-reversal $\mathcal{T} = \eta_z \mathcal{K} $ symmetries, where $\mathcal{K}$ denotes complex conjugation.
The eigenenergies of the Hamiltonian Eq.~\eqref{eq:MFHam} can be obtained analytically: 
\begin{align} \label{eq:MFEigval}
\begin{split}
& E^2 (k_x,k_y) =   \epsilon_k^2 + V_z^2 + \Delta^2 + \alpha^2 \sin^2 k_x + \\
& \beta_1^2 + \beta_2^2 - 2\beta_1 \beta_2 \cos k_y \pm \\
& \sqrt{4 V_z^2 (\epsilon_k^2 + \Delta^2) + 4 \epsilon_k^2 (\alpha^2 \sin^2 k_x + 
 \beta_1^2 + \beta_2^2 -2\beta_1 \beta_2 \cos k_y)}, 
\end{split}
\end{align}
where $\epsilon_k = 2t_x(1- \cos{k_x})  -\mu$. When $\beta_1 = \beta_2$, $\mu=0$ and $V_z=\pm \Delta$, the model admits gapless excitations at $\mathbf{k} = (0,0)$ while for  $\beta_1 = -\beta_2$, $\mu=4 t_x$ and $V_z=\pm \Delta$, these bulk excitations are at  $\mathbf{k}= (\pi,\pi)$.

\begin{figure}[tb]
\begin{center}
\includegraphics[width=0.6\columnwidth]{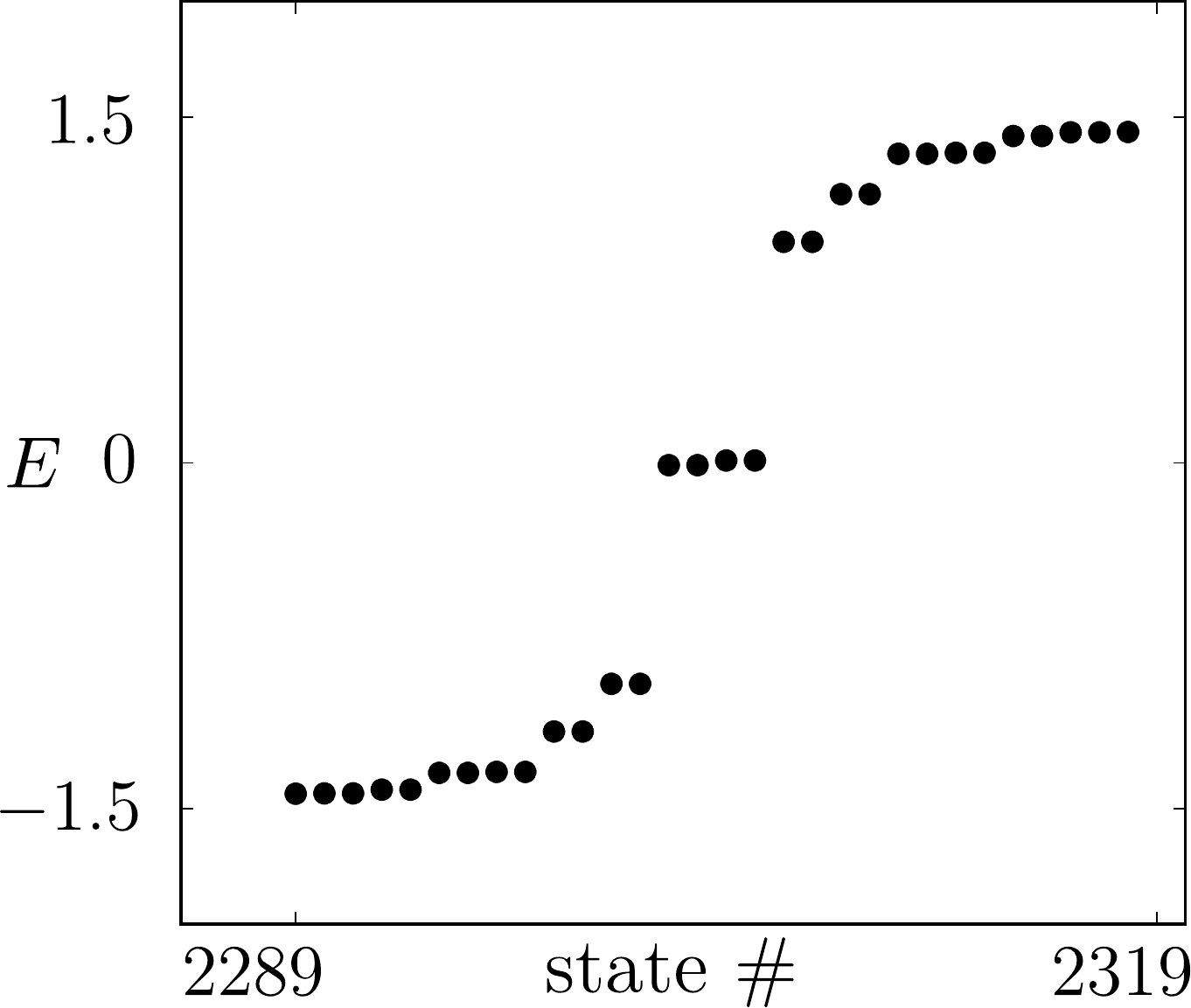}
\end{center}
\caption{The energy spectrum for a model given by Eq.~\eqref{eq:MFHam} for a system of $24 \times 24$ unit cells. The parameters are given in the text. The bulk spectrum is gapped, and there are four zero-energy modes, one localized at each corner.}
\label{fig:Espectra}
\end{figure}

Throughout the following, we set, unless otherwise specified, $\mu=-0.9$, $t_x=1.6$, $V_z=2.7$, $\Delta=1.6$, $\alpha=3.7$, $\beta_{avg}=3.5$ and $\beta_{dim}=2.7$, so that $\beta_1=0.8$ and $\beta_2=6.2$. 
For these parameters, the bulk spectrum is gapped and we find one zero-energy mode localized at every corner, as shown in Fig.~\ref{fig:Espectra}.

\begin{figure}[tb]
\begin{center}
\includegraphics[width=\columnwidth]{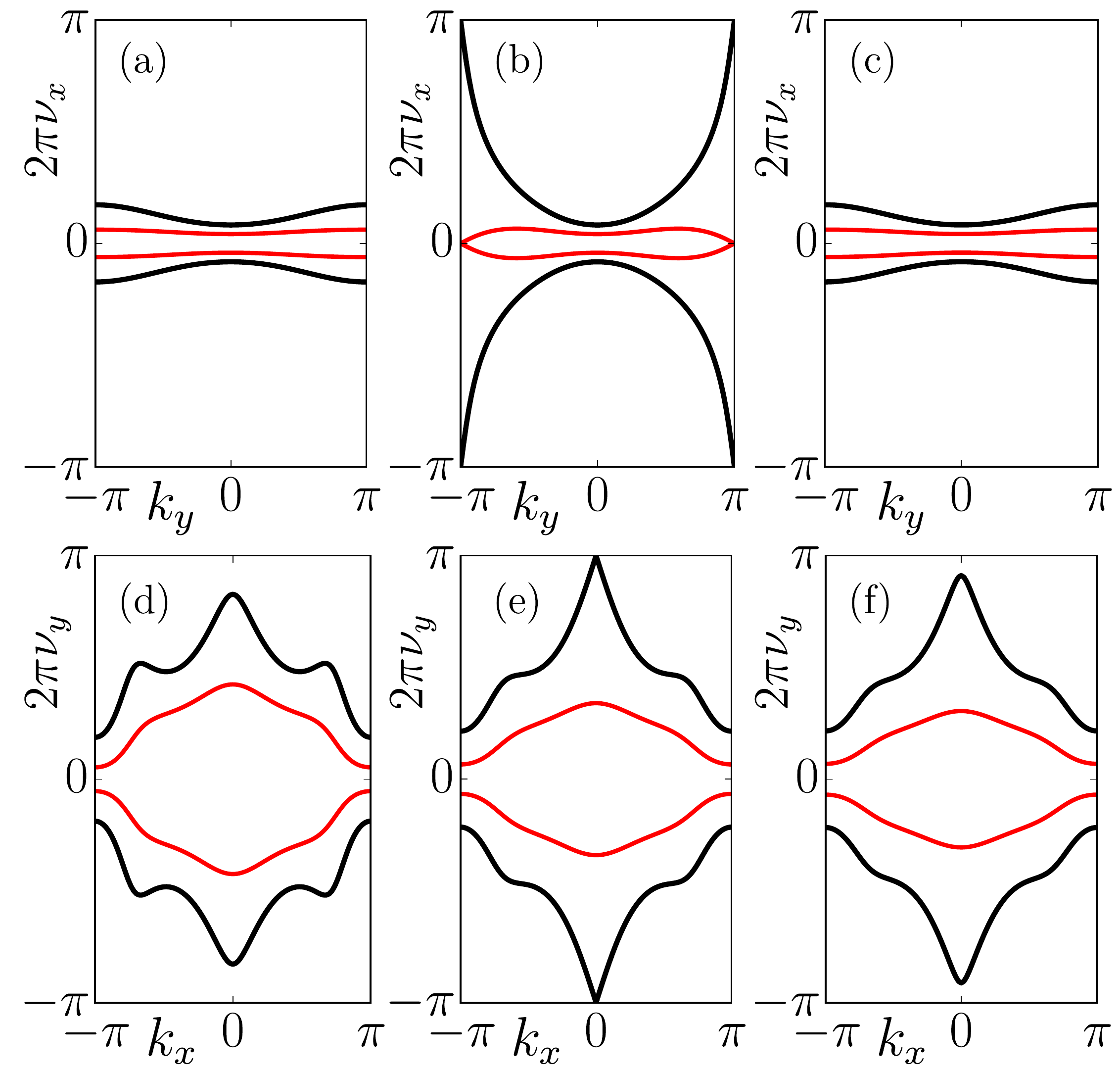}
\end{center}
\caption{(a-c) Wannier bands in the case of an anomalous HOTI obtained by calculating the Wilson loop operator in the $k_x$-direction:  (a) in a HOTI phase: $\beta_{avg}=3.5$ and $\beta_{dim} = 2.7$, 
 (b) at the transition line that corresponds to a weak topological insulator phase: $\beta_{avg}=3.5$ and $\beta_{dim} =0$ and (c) in a phase described by $p^{\nu}=((1/2,0),(0,0))$, when  $\beta_{avg}=3.5$ and $\beta_{dim} = -2.7$. 
(d-f) Wannier bands obtained by calculating the Wilson loop operator in the $k_y$-direction: (d) in a HOTI phase where $V_z > \sqrt{\Delta^2 + \mu'^{2}}$ is valid, 
(e) at the transition line (when  $V_z = \sqrt{\Delta^2 + \mu'^{2}} $; $\mu'$ is given in the text)  and (f) in a phase described by $p^{\nu}=((0,0),(1/2,1/2))$ corresponding to $V_z < \sqrt{\Delta^2 + \mu'^{2}} $ ($\Delta=3.2$). The initial point of the Wilson loop is at $(k_x,k_y)=(\pi,\pi)$  and  the number of k-points is $101$.} 
\label{fig:Transition}
\end{figure}

Since this model is not rotationally invariant, the Wannier spectra (see Fig.~\ref{fig:Transition}) are different in the two directions. The presence of four Wannier bands, two per each Wannier sector, is the consequence of four occupied bands. 
The diagonalization of the Wilson loop $\mathcal{W}_{x,(\pi,\pi)}$ reveals the spectrum adiabatically connected to the physical y edge Hamiltonian. If the topological nanowires are in the non-trivial regime, we expect that the gap closing will happen once the spin-orbit coupling strengths $\beta_1$ and $\beta_2$ become equal, as shown in Fig.~\ref{fig:Transition}b.  On this panel, we see that the gap closing happens for both bands in the Wannier sector. This indicates that all the Wannier bands are topologically non-trivial in a HOTI phase and that expectation is confirmed by performing integrals over Berry connection for each Wannier band separately. The resulting polarization of each band is $1/2$, i.e. $(p_y^{\nu^{\pm}_{x},1},p_y^{\nu^{\pm}_{x},2})= (1/2,1/2)$ (black and red Wannier bands are denoted by superscripts $1$ and $2$, respectively). However, the Wannier sector polarization $p_y^{\nu_x^{\pm}}$ is always the sum of bands' invariants in a particular sector thus giving a trivial result.
Note that once the intracell and intercell pairings between the wires $\beta_1$ and $\beta_2$ become equal, the system will be in a weak topological insulator phase with two, counter-propagating edge modes along the y edge. 

In the same manner, the Wannier bands along $k_x$ are adiabatically connected to the x edge Hamiltonian. It can be expected that the physical x edge Hamiltonian resembles the topological nanowire Hamiltonian with quantities that are renormalized under the spin-orbit coupling. Namely, in the case of a single nanowire, the gap closing occurs when $V_z = \sqrt{\mu^2 + \Delta^2}$ \cite{Oreg2010,Lutchyn2010}. This condition is affected once the wires are coupled since the chemical potential is renormalized. However, we indeed see that the gap closing (Fig.~\ref{fig:Transition}e) for the Wannier bands occurs when $V_z = \sqrt{\mu'^{2} + \Delta^2}$ where $\mu'$ is the renormalized chemical potential. We numerically find that $\mu' \approx 0.46$ for the previously defined parameters. The gap closing at $\pi$ happens for only one Wannier band per Wannier sector, suggesting that $(p_x^{\nu^{\pm}_{y},1},p_x^{\nu^{\pm}_{y},2})= (1/2,0)$ in the TQI phase, so that the Wannier sector polarization $p_x^{\nu_y^{\pm}}$ is nontrivial. This expectation is indeed confirmed when computing the polarizations of each of the bands, as explained in the main text.

\begin{figure}[tb]
\begin{center}
\includegraphics[width=\columnwidth]{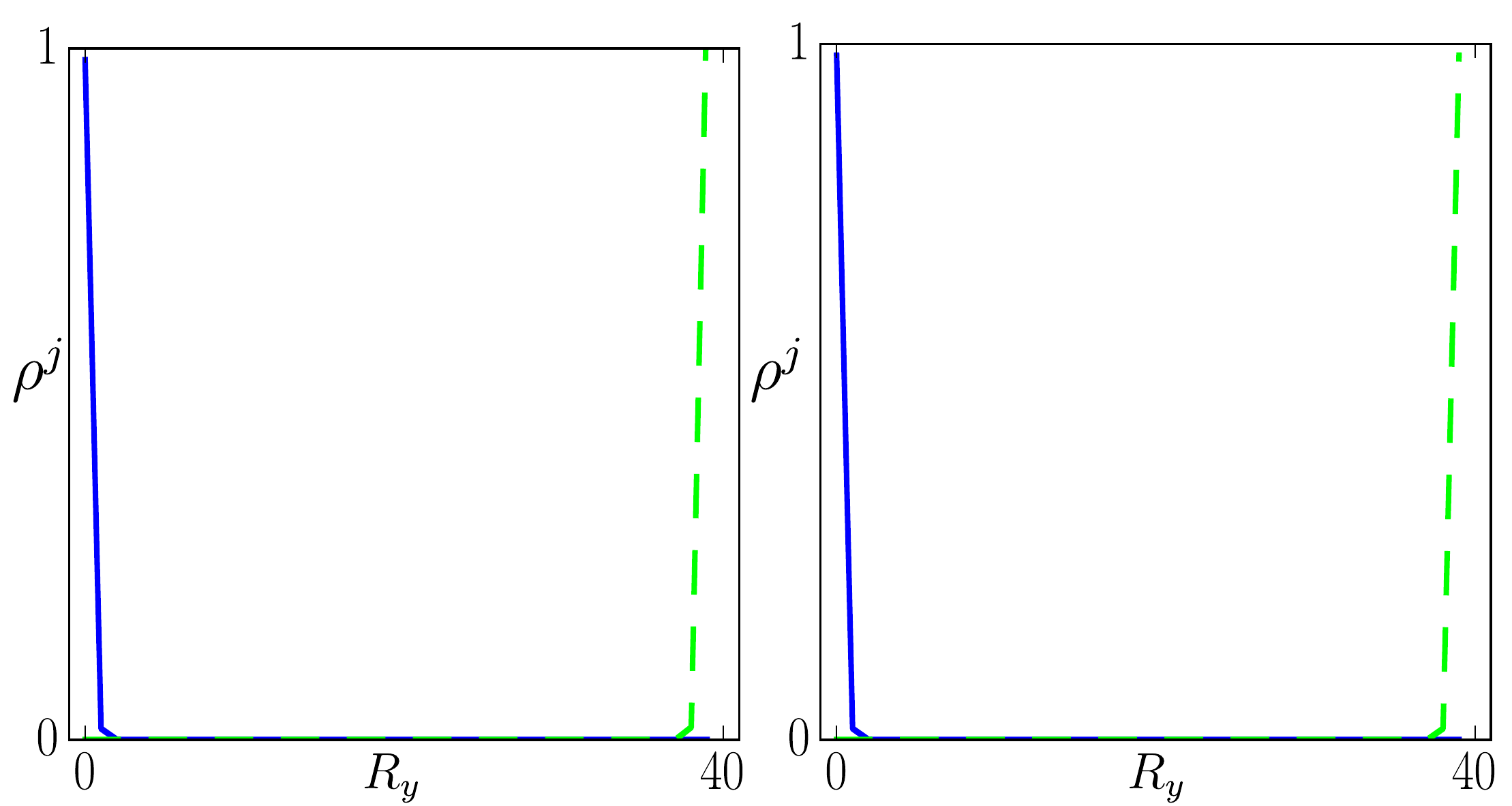}
\end{center}
\caption{The probability distribution of Wannier states as a function of unit cell index corresponding to $0$-modes (left panel) and to $\pi$-modes (right panel) in the case of an anomalous HOTI. Here, we consider a strip with PB in the x direction. In the left panel, we see two $0$-modes (solid blue and dashed green lines) that are peaked on different edges and whose overlap is exponentially small.  The same holds for $\pi$-modes. We use a momentum grid of $51$ k-points and the system size in the direction of open boundaries is $40$ unit cells.}
\label{fig:ProbDist}
\end{figure}

In the strip geometry with periodic boundaries (PB) in the x direction, the fact that all Wannier bands calculated from the Wilson loop operator in the $k_x$ direction (Fig.~\ref{fig:Transition}a) are non-trivial results in the presence of two kinds of modes, $0$- and $\pi$-modes. We find that both $0$- and $\pi$-modes are peaked at the open boundaries of the system (Fig.~\ref{fig:ProbDist}). However, the presence of $0$-eigenphases in the spectrum does not influence the tangential edge polarization since it is a combination of the Wilson loop eigenvalues and the probability distribution of each mode (see Appendix 2). As already shown in the main text, the integrated tangential edge polarization remains quantized in both directions and has the same magnitude as the corner charge. 

\section{Appendix 2: Details of calculations in various geometries}

\subsection{Bulk}

In the translationally invariant system, polarization results from a displacement of electrons with respect to the positions of positive charges in the unit cell. By matching the center of unit cell to the center of positive charge, the only relevant contribution to the polarization comes from negative charges. To calculate the polarization, one needs to know the positions of all electrons in the system and this can be obtained from diagonalization of the projected position operator onto the occupied subspace \cite{Resta1998}.
The projection operator onto the occupied subspace is
\begin{equation*}
P^\text{occ} = \sum_{n=1}^{N_\text{occ}} \sum_{\bold{k}} \gamma_{n,\bold{k}}^{\dagger} \ket{0} \bra{0} \gamma_{n,\bold{k}}^{\phantom{\dagger}},
\end{equation*}
where $N_\text{occ}$ denotes the number of occupied bands and $\bold{k}$ is the momentum. The latter is discretized in the 2D BZ using steps $(2\pi/N, 2\pi/N)$, such that there are $N^2$ k-points in total. Furthermore, $\ket{0}$ denotes the vacuum, $\gamma_{n,\bold{k}}^{\dagger} = \sum_{\alpha=1}^{N_{orb}} [u_{\bold{k}}^{n}]^{\alpha}  c_{\bold{k},\alpha}^{\dagger} $ where $c^{\dagger}_{\bold{k},\alpha}$ is the creation operator and $[u_{\bold{k}}^{n}]^{\alpha}$ are the elements of the 
the n-th eigenstate of the Hamiltonian, $\ket{u^n_{\bold{k}}}$. The index $\alpha=1,...,N_{orb}$, with $N_{orb}$ the total number of orbitals.

This operator is used to form a projected position operator: 
\begin{equation*}
P^\text{occ} \hat{x} P^\text{occ} = \sum_{m,n=1}^{N_\text{occ}} \sum_{\bold{k}} \gamma_{m,\bold{k} + \Delta_{ \bold{k} } }^{\dagger} \ket{0} \braket{u_{ \bold{k}+\Delta_{ \bold{k} } }^m \; | \; u_{ \bold{k} }^n}  \bra{0} \gamma_{n, \bold{k} }^{\phantom{\dagger}},
\end{equation*}
where $\Delta_{\bold{k}}=(2\pi/N,0)$. The projected position operator is a matrix of dimensions $N_{orb} \times N_{orb}$ and it is diagonal in $k_y$. By successive application of the eigenvalue equation $P^\text{occ} \hat{x} P^\text{occ} \ket{\Psi^j} = E^j \ket{\Psi^j}$ (where $\ket{\Psi^j}$ are Wannier states), we can obtain the Wilson loop operator in the discrete limit 
\begin{equation}\label{eq:WLx}
\mathcal{\tilde{W}}_{x,\mathbf{k}_0} = G_{x,\mathbf{k}_{N-1}} ... G_{x,\mathbf{k}_2}  G_{x,\mathbf{k}_1}  G_{x,\mathbf{k}_0},
\end{equation}
where $\mathbf{k}_0= (k_{0,x},k_{0,y})$ is the initial point of the Wilson loop operator and $\mathbf{k}_l = \mathbf{k}_0 + l \Delta_{\mathbf{k}}$ ($l=0,...,N-1$). The element $G_{x,\mathbf{k}_l}$ is a matrix of of dimensions $N_\text{occ} \times N_\text{occ}$ defined as
\begin{equation}
[G_{x,\mathbf{k}_l}]^{mn} =  \braket{u_{{\mathbf{k}_l} + \Delta_{\mathbf{k}}}^m \; | \; u_{\mathbf{k}_l}^n}.
\end{equation} 
The matrix $G$ is not a unitary matrix due to the discretization of momentum $\bold{k}$ \cite{Benalcazar2017, Benalcazar2017a}. However, it can be unitarized using the singular value decomposition (SVD) at every momentum point. Namely, from $G_{x,\bold{k}_l} = U D V^{\dagger}$ where $D$ is a diagonal matrix with elements smaller than one, a new unitary matrix $F_{x,\bold{k}_l}$ is defined as $F_{x,\bold{k}_l}=UV^{\dagger}$. This new matrix $F_{x,\bold{k}_l}$ replaces $G_{x,\bold{k}_l}$ at every momentum point such that a unitary Wilson loop operator in the discrete limit becomes
\begin{equation}\label{eq:WLx1}
\mathcal{W}_{x,\mathbf{k}_0} = F_{x,\mathbf{k}_{N-1}} ... F_{x,\mathbf{k}_2}  F_{x,\mathbf{k}_1}  F_{x,\mathbf{k}_0}.
\end{equation}

This operator is in the thermodynamic limit defined
as the path-ordered integral (denoted $\overline \exp$) over the Berry connection \cite{ Wilczek1984, Berry1984, Benalcazar2017a}
\begin{equation}\label{eq:WL0}
\mathcal{W}_{x,\bold{k}_0} =\overline \exp \left(- i \int_{\bold{k}_0}^{\bold{k}_0 + 2\pi} d \bold{k} \; \mathcal{A}_{\bold{k}} \right),
\end{equation}
where $  [\mathcal{A}_{ \bold{k} }]^{m n } =- i \bra{u^m_{\bold{k}}} \nabla_{\bold{k}} \ket{u^n_{\bold{k}}}$ is a Berry connection of the occupied eigenstates of the Hamiltonian. The Wilson loop operator describes the parallel transport of eigenstates along one direction ($k_x$ in this case) over the whole BZ. 
Moreover, it is a unitary operator, a fact which can be straightforwardly seen from Eq.~\eqref{eq:WLx1}, which contains a product of unitaries. Its eigenvalue equation reads
\begin{equation}\label{WLoopEig}
\mathcal{W}_{x,\bold{k}_0} \ket{\nu_{x,\bold{k}_0}^{\pm,r}} = \exp \left[ i 2 \pi \nu_x^{\pm,r} (k_y) \right] \ket{\nu_{x,\bold{k}_0}^{\pm,r}},
\end{equation} 
where $\pm$ denotes the Wannier sector and $r=1,...,N_\text{occ}/2$, while $\nu_x^{\pm,r} (k_y)$ are the eigenvalues that correspond to Wannier centers \cite{Fidkowski2011}. In the eigenvalue equation, we have separated $N_\text{occ}$ eigenvalues of a matrix with dimensions $N_\text{occ} \times N_\text{occ}$ into two Wannier sectors. Due to the lattice symmetries \cite{Benalcazar2017,Benalcazar2017a}, there are exactly $N_\text{occ}/2$ eigenvalues that are positive and belong to a positive Wannier sector and vice versa. These eigenvalues are plotted on Figs.~\ref{fig:SSHTransition} and \ref{fig:Transition} in the case of a 2D SSH model and a model of coupled nanowires, respectively.

The Wilson loop eigenstates $\ket{\nu_{x,\bold{k}}^{\pm,r}}$ with their components $ [\nu_{x,\mathbf{k}}^{\pm,r}]^n$  are used to form Wannier states, the eigenstates of the projected position operator, as
\begin{equation}\label{eq:wannier_subspace1}
\ket{w_{x,\mathbf{k}}^{\pm,r}}  = \sum_{n=1}^{N_\text{occ}} \ket{u_{\mathbf{k}}^n} [\nu_{x,\mathbf{k}}^{\pm,r}]^n .
\end{equation}

Note that the Wilson loop eigenstates (and consequently Wannier states) depend on the initial point $\bold{k}$ of the Wilson loop operator, unlike its eigenvalues that depend only on the remaining momentum ($k_y$ in this case). 
These states are used to calculate the Wannier sector polarization as: 
\begin{equation}\label{eq:FinalPol}
p_y^{\nu_x^\pm} = - \frac{1}{(2\pi)^2} \int_{BZ} {\rm Tr}\left[ \tilde{\mathcal{A}}_{y,\mathbf{k}}^{\nu_x^\pm} \right] d^2\mathbf{k},
\end{equation}
where $[\tilde{\mathcal{A}}_{y,{\bf k}}^{\nu_x^\pm}]^{mn} = - i \bra{w_{x,{\bf k}}^{\pm,m}} \partial_{k_y} \ket{w_{x,{\bf k}}^{\pm,n}}$ are the matrix elements of the $(N_\text{occ}/2)\times (N_\text{occ}/2)$ Berry connection of the $\pm$ Wannier subspace.
This is a is a $\mathbb{Z}_2$ topological index, $p_y^{\nu_x^\pm}\in\{0,1/2\}$.

Since the Wilson loop operator can only be computed numerically, it is necessary to have numerical eigenstates continuous over the whole integration contour. For a non-degenerate state, the numerical subroutine returns an eigenvector with a random phase factor at every momentum point. This phase factor makes the Berry phase discontinuous and therefore has to be eliminated. This problem can be resolved by isolating the component of an eigenvector that has the largest overall absolute value in the BZ and making such a component real. Note that the use of such a component is done in order to avoid possible errors in computation while treating small numbers. 

However, for the models studied in the main text, the problem is more complex. The energy levels are doubly degenerate and the calculated eigenvectors are at each momentum point given in a different basis. This basis $(\ket{u_1'} \;  \ket{u_2'} )$ is a random combination of the continuous eigenvectors $(\ket{u_1} \;  \ket{u_2} )$ defined over the BZ. This is represented as
\begin{equation}
(\ket{u_1'} \;  \ket{u_2'} ) = U (\ket{u_1} \;  \ket{u_2}),
\end{equation}
where $U$ is a unitary matrix defined as: 
\begin{equation} \label{U}
U= e^{i \psi} 
 \left( {
\begin{array}{cc}
e^{i \psi_1} \cos{\theta} & e^{i \psi_2} \sin{\theta} \\
- e^{-i \psi_2} \sin{\theta} & e^{-i \psi_1} \cos{\theta}
\end{array} } \right  ), 
\end{equation}
and $\psi$, $\psi_1$, $\psi_2$ and $\theta$ are quantities that should be determined at each momentum point. In the code we have provided, there is a general and stable basis fixing procedure in the case of doubly-degenerate states. This fixing procedure again implements the idea of making the elements of the two eigenvectors with the largest overall amplitudes in the BZ real, and then rotating them until they become zero. We have then "disentangled" two vectors ($\ket{u_1}$ and $\ket{u_2}$) and the final step would be to make both vectors smooth throughout the BZ. This is done in a fashion similar to the case of a non-degenerate state, i.e., by making the elements with largest absolute values real in both vectors. 

\subsection{Strip geometry}

As already discussed in the main text, the strip geometry reveals the presence of edge modes in the Wannier spectrum. These edge modes are in one-to-one correspondence to non-trivial Wannier bands in the bulk. They are also responsible for the quantized tangential edge polarization. In order to determine all these quantities, we need to work with tight-binding Hamiltonians represented as follows. 

The Hamiltonian Eq.~\eqref{eq:BBHHam} in the strip geometry with PB in the $k_x$ direction becomes:
\begin{align*} 
\begin{split}
& H({k_x,N_y}) {} =\sum_{j=1}^{N_y} \Big\{ \Psi_j^{\dagger} [ (\gamma_x  + \lambda_x \cos{k_x}) \tau_x \sigma_0 - \lambda_x \sin{k_x} \tau_y \sigma_z \\
& -  \gamma_y  \tau_y \sigma_y] \Psi_j -\frac{1}{2} \{ \Psi_j^{\dagger} [  \lambda_y \tau_y \sigma_y - i \lambda_y \tau_y \sigma_x ] \Psi_{j+1} + \rm{h.c.} \} \Big\},
\end{split}
\end{align*}
where the basis is $\Psi_j^{\dagger} = ( \rm{c}^{\dagger}_{j,\rm{A},1}, \rm{c}^{\dagger}_{j,\rm{A},2},  \rm{c}^{\dagger}_{j,\rm{B},1}, \rm{c}^{\dagger}_{j,\rm{B},2})$. A and B denote different sublattice degree of freedom, $1,2$ different sites belonging to the same sublattice, and $j$ is an integer labeling the unit cells. 

Analogously, the Hamiltonian Eq.~\eqref{eq:MFHam} becomes, in the case of a strip with finite boundaries in the y direction:
\begin{align*} 
\begin{split}
& H(k_x,N_y) {} =\sum_{j=1}^{N_y} \Big \{  \Psi_j^{\dagger} \{ [2t_x(1- \cos{k_x})  -\mu]\tau_z \sigma_0 \eta_0  + \\ 
&   V_z \tau_0 \sigma_z \eta_0 +  \Delta \tau_x \sigma_0 \eta_0 + \alpha \sin{k_x} \tau_z \sigma_y \eta_0  -  \\
& \beta_1 \tau_z \sigma_x \eta_y \} \Psi_{j} + \frac{1}{2} \{ \Psi_j^{\dagger} [i \beta_2  \tau_z \sigma_x \eta_x + \beta_2  \tau_z \sigma_x \eta_y ]\Psi_{j+1} +\rm{h.c.} \} \Big \},
\end{split}
\end{align*}
while, in the case of a strip with finite boundaries in the x direction:
\begin{align*} \label{MF_TBy}
\begin{split}
& H(N_x,k_y) {} =\sum_{j=1}^{N_x} \Big \{  \Psi_j^{\dagger} \{ [2t_x  -\mu]\tau_z \sigma_0 \eta_0  +  V_z \tau_0 \sigma_z \eta_0 + \\ 
&   \Delta \tau_x \sigma_0 \eta_0 + \beta_1 \tau_z \sigma_x \eta_y - \beta_2 \sin{k_y} \tau_z \sigma_x \eta_x + \\
& \beta_2 \cos{k_y} \tau_z \sigma_x \eta_y \} \Psi_j -  \frac{1}{2} \{\Psi_j^{\dagger} [ 2 t_x - i \alpha   \tau_z \sigma_y \eta_0 ] \Psi_{j+1}^{\dagger} +\rm{h.c.} \} \Big \}, 
\end{split}
\end{align*}
where  
\[\Psi_j^{\dagger} = ( \rm{c}^{\dagger}_{j,\uparrow,1}, \rm{c}^{\dagger}_{j,\uparrow,2},  \rm{c}^{\dagger}_{j,\downarrow,1}, \rm{c}^{\dagger}_{j,\downarrow,2},\rm{c}_{j,\downarrow,1}, \rm{c}_{j,\downarrow,2},  -\rm{c}_{j,\uparrow,1}, -\rm{c}_{j,\uparrow,2}). \]
Here, $\Psi_j^{\dagger}$ is given in the basis of particle-hole $\times$ spin $\times$ wire spaces.

In order to obtain a finite edge polarization, both mirror symmetries should be broken while the inversion symmetry is preserved. Both perturbations introduced in the main text for this purpose are onsite terms, which implies that they remain the same for all geometries. Specifically, we consider $\delta \tau_z \sigma_0$ for a 2D SSH model and $\delta \tau_y \sigma_y \eta_z$ for a model of anomalous HOTI.

Obtaining the Wannier spectrum and tangential polarization requires the diagonalization of a Wilson loop operator, which in a strip geometry can only be calculated in the direction with PB. Let us consider the strip with open boundaries in the y direction. Then, the Wilson loop calculation is performed in the discrete limit ($k_x$ from ${-\pi}$ to ${\pi}$ with steps $\Delta_{k_x}=2\pi /N$ where $N$ is the total number of unit cells in the direction with PB) as in Eq.~\eqref{eq:WLx} \cite{Benalcazar2017a}. However, the Wilson loop operator depends also on the real space index and is a matrix of dimensions $(N_\text{occ} \times N_y) \times (N_\text{occ} \times N_y)$. The Wilson loop elements are:
\begin{equation}
[G_{k_x}]^{mn} = \braket{u^m_{k_x+\Delta_{k_x}} \; | \; u_{k_x}^n},
\end{equation}
where $\ket{ u_{k_x}^n}$ is the eigenstate of the Hamiltonian with components $[u_{k_x}^n]^{R_y,\alpha}$, with $\alpha=1,...,N_{orb}$ and $R_y=1,...,N_y$ labels the unit cell index in the y direction. 
In this Wilson loop calculation, we effectively treat a 2D system as a quasi-1D system, with the degrees of freedom associated to the y direction corresponding to the degrees of freedom of the unit cell.

Analogously to the case of the bulk, the matrices $G_{k_x}$ have to be unitarized using the SVD after which we obtain unitary matrices $F_{k_x}$ and calculate the Wilson loop operator $\mathcal{W}_{x,k_x}$ as in Eq.~\eqref{eq:WLx1}. Its
eigenvalue equation is
\begin{equation}
\mathcal{W}_{x,k_x} \ket{\nu_{x,k_x}^{j}} = \exp \left[ i 2 \pi \nu_x^{j} \right] \ket{\nu_{x,k_x}^{j}},
\end{equation}
where $k_x$ denotes the initial point of the Wilson loop, $j=1,...,N_\text{occ} \times N_y$, and $\nu_x^{j}$ are eigenphases. 
The Wilson loop eigenstates $\ket{\nu_{x,k_x}^{j}}$ are used to calculate the Wannier states  
\begin{equation}\label{eq:wannier_subspace2}
\ket{w_{x,k_x}^{j}}  = \sum_{n=1}^{N_\text{occ} \times N_y} \ket{u_{k_x}^n} [\nu_{x,k_x}^{j}]^n,
\end{equation}
and the hybrid Wannier functions
\begin{equation}
\ket{\Psi^j_{R_x} }= \frac{1}{\sqrt{N}} \sum_{n=1}^{N_\text{occ} \times N_y} \sum_{k_x}  [\nu_{k_{x}}^{j}]^n e^{-i k_x R_x} \gamma_{n,k_x}^{\dagger} \ket{0},
\end{equation}
where $R_x = 1,...,N$ labels the unit cell position,  [$\nu_{k_{x}}^{j}]^n$ is n-th component of the j-th eigenstate of the Wilson loop operator and $\gamma^{\dagger}_{n,k_x} = \sum_{R_y,\alpha} [u_{k_x}^{n} ]^{R_y,\alpha} c^{\dagger}_{k_x,R_y,\alpha}$  where $ c^{\dagger}_{k_x,R_y,\alpha}$ is the particle creation operator. 

The Wannier functions are necessary to define the probability density $\rho^{j,R_x} (R_y) \equiv \rho^j (R_y)$:
\begin{align}
\begin{split}
\rho^j (R_y) =&  \braket{\Psi^j_{R_x} | \Psi^j_{R_x}} =\sum_{k_x} \braket{w_{x,k_x}^{j} | w_{x,k_x}^{j}} = \\
&\frac{1}{N} \sum_{k_x,\alpha} | [u_{k_x}^n ]^{R_y,\alpha} [\nu_{k_{x}}^{j}]^n |^2.
\end{split}
\end{align}
These hybrid Wannier functions are used to determine which modes are localized on the open edges of the system, as plotted for the case of $0$- and $\pi$-modes (Fig.~\ref{fig:ProbDist}) in Appendix 1. 

Finally, the tangential polarization is given as
\begin{equation}\label{tan_pol}
p_x (R_y) =  \sum_{j=1}^{N_\text{occ} \times N_y} \rho^j (R_y) \nu_x^j.
\end{equation}

\subsection{Finite-size systems}
At the end, we briefly discuss the calculation of the charge density distribution. The presence of protected corner charges can only be detected in a finite-size system once the degeneracy of four corner modes is broken by a perturbation that preserves the inversion symmetry but breaks both mirrors. 

Let integers $R_x$ and $R_y$ index the positions of unit cells in a system of dimensions $N_x \times N_y$ and $m=1,...,N_\text{occ}$. We can then label the elements of the occupied eigenstates $\ket{u_m}$ according to the unit cell index: $[ u_m ]^{R_x,R_y}$. With this indexing, the probability distribution takes the form $\rho(R_x,R_y) = \sum_m |[ u_m ]^{R_x,R_y}|^2$. The charge density distribution is then simply $\rho^{\rm{charge}}_{R_x,R_y} = e \rho_{R_x,R_y}$.

\section{Appendix 3: Proof that the topological invariant is quantized}

In this appendix, we prove that the polarization of each individual Wannier band,
\begin{equation}\label{eq:PolWy1}
p_y^{\nu_x^\pm, r} =  \frac{i}{(2\pi)^2} \int_{BZ} \bra{w_{x,{\bf k}}^{\pm,r}} \partial_{k_y} \ket{w_{x,{\bf k}}^{\pm,r}} d^2\mathbf{k}
\end{equation}
is quantized, thus resulting in a quantized bulk quadrupole moment. The proof is based on equations derived in Ref.~\cite{Benalcazar2017a}, Appendix D, section 2.b. There, it has been proven that crystalline symmetries are essential for a a quantized Wannier sector polarization [Eq.~\eqref{eq:FinalPol}]. We find that this is also the case in the model of an anomalous higher-order topological insulator, \emph{i.e.} crystalline symmetries are required for an index Eq.~\eqref{eq:PolWy1} to take quantized values.

Let us start from Eq. (D.41) in Ref.~\cite{Benalcazar2017a}:
\begin{equation} \label{Wstates}
B_{{\cal M}_y \bold{k}} \ket{\nu_{x, \bold{k}}^i} = \ket{\nu_{x,  \mathcal{M}_y \bold{k}}^j} \alpha_{\mathcal{M}_y \bold{k}}^{ji}, 
\end{equation} 
where summation up to $N_\text{occ}$ is implied over repeated indexes, $ {\mathcal{M}}_y \bold{k} \stackrel{\text{def.}}{=} (k_x,-k_y)$, $B_{\mathcal{M}_y \bold{k}}$ is a unitary sewing matrix of Hamiltonian eigenstates $\ket{u_{\bold{k}}^n}$, defined as 
\begin{equation}
B_{\mathcal{M}_y \bold{k}}^{mn} = \bra{u_{ \mathcal{M}_y \bold{k}}^m} \mathcal{M}_y \ket{u_{\bold{k}}^n}
\end{equation}
and $\alpha_{\mathcal{M}_y \bold{k}}$ is a unitary sewing matrix of Wilson loop eigenstates $\ket{\nu_{x,\bold{k}}^i}$
\begin{equation} \label{alpha}
\alpha_{\mathcal{M}_y \bold{k}}^{ji} = \bra{ \nu_{x, \mathcal{M}_y \bold{k}}^j } B_{ \mathcal{M}_y \bold{k}} \ket{\nu_{x, \bold{k}}^i}.
\end{equation} 
The above relations, together with the effect of mirror symmetry on the Hamiltonian eigenstates, ${\mathcal{M}}_y \ket{u_{\bold{k}}^n} = \ket{u_{\mathcal{M}_y \bold{k}}^m} B_{\mathcal{M}_y \bold{k}}^{mn}$ allows one to determine the transformation law of the Wannier states $\ket{w_{x,\mathbf{k}}^{\pm,r}}$ [Eq.~\eqref{eq:wannier_subspace1}] due to the mirror symmetry in the y-direction:
\begin{equation}\label{eq:wannier_state_mirror}
\ket{w_{x, \mathcal{M}_y \mathbf{k}}^{\pm,r}} =  {\mathcal{M}}_y   \ket{w_{x, \mathbf{k}}^{\pm,s}} \alpha_{\mathcal{M}_y \bold{k}}^{sr}.
\end{equation}

Crucially, as shown in Ref.~\cite{Benalcazar2017a}, the matrix elements $\alpha_{\mathcal{M}_y \bold{k}}^{sr}\neq0$ only when the Wannier bands obey $\nu_x^s(k_y)=\nu_x^r(-k_y)$. Since the Wannier bands of our model do not cross, and since they are positioned symmetrically around $k_y=0$ due to the ${\cal M}_y$ symmetry, this implies that the unitary matrix $\alpha_{\mathcal{M}_y \bold{k}}$ is just a diagonal matrix of phases. Therefore, $\alpha_{\mathcal{M}_y \bold{k}}^{sr} = \delta_{rs} e^{i \phi^r_{\bold{k}}}$ and $ \phi^r_{\bold{k}}$ is an arbitrary phase that is periodic in momentum such that $ \phi^r_{(k_x,k_y)} =  \phi^r_{(k_x,k_y+2\pi)} + 2\pi n$ (n is an integer number).

Using Eq.~\eqref{eq:wannier_state_mirror}, the Berry connection of a given Wannier state, $\tilde{\mathcal{A}}_{y,\mathbf{k}}^{\nu_x^\pm, r } = - i \bra{w_{x,{\bf k}}^{\pm,r}} \partial_{k_y} \ket{w_{x,{\bf k}}^{\pm,r}}$ transforms as
 
\begin{align}
\begin{split}
\tilde{\mathcal{A}}_{y, \mathcal{M}_y\mathbf{k}}^{\nu_x^\pm, r } & = -i \bra{w_{x,{\mathcal{M}_y \bf k}}^{\pm,r}} \partial_{\mathcal{M}_y k_y} \ket{w_{x,{\mathcal{M}_y \bf k}}^{\pm,r}} = \\
& = - \tilde{\mathcal{A}}_{y,\mathbf{k}}^{\nu_x^\pm, r } - \partial_{k_y} \phi^r_{\bold{k}}
\end{split}
\end{align}
Thus, the polarization of a Wannier band obeys
\begin{align}
\begin{split}
p_y^{\nu_x^\pm, r} = & -  \frac{1}{(2\pi)^2} \int_{-\pi}^{\pi} d k_x \int_{-\pi}^{\pi} d k_y  \tilde{\mathcal{A}}_{y,\mathbf{k}}^{\nu_x^\pm, r } \\
= & \frac{1}{(2\pi)^2} \int_{-\pi}^{\pi} d k_x \int_{-\pi}^{\pi} d k_y \tilde{\mathcal{A}}_{y, \mathcal{M}_y\mathbf{k}}^{\nu_x^\pm, r } + \\
&   \frac{1}{(2\pi)^2} \int_{-\pi}^{\pi} d k_x \int_{-\pi}^{\pi} d k_y  \partial_{k_y} \phi^r_{\bold{k}}  \\
 =&  - p_y^{\nu_x^\pm, r} + n
\end{split}
\end{align}
The last equation implies $p_y^{\nu_x^\pm, r} = -p_y^{\nu_x^\pm, r} \text{ mod } 1$, such that $p_y^{\nu_x^\pm, r} = 0$ or $1/2$. In a similar fashion, the mirror symmetry in the x direction ${\cal M}_x$ is responsible for quantizing $p_x^{\nu_y^\pm, r}$, since the Wannier bands of $\mathcal{W}_{y,\bold{k}}$ are also non-degenerate and symmetric in momentum.
Thus, the bulk quadrupole moment $q_{xy}$ obtained as a combination of the Wannier bands polarizations takes discrete values of $0$ for a trivial phase and $\frac{1}{2}$ for a quadrupole phase. 

\bigskip
\section{Appendix 4: Proof that the corner-localized zero-energy state is the simultaneous edge state of both low-energy edge Hamiltonians}

The corner mode should be the simultaneous eigenstate of both edge Hamiltonians, for instance the topological nanowire Hamiltonian on the top edge and the Kitaev chain Hamiltonian on the right edge. To verify that this is the case, it is sufficient to consider the continuum version of Hamiltonian Eq.~\eqref{eq:MFHam}, when $\mu=0$: 
\begin{align*} 
\begin{split}
H(\mathbf{k}) {} = & V_z \tau_0 \sigma_z \eta_0 +  \Delta \tau_x \sigma_0 \eta_0 + \alpha k_x \tau_z \sigma_y \eta_0  -  \\
& \beta_1 \tau_z \sigma_x \eta_y - \beta_2 k_y \tau_z \sigma_x \eta_x + \beta_2  \tau_z \sigma_x \eta_y.
\end{split}
\end{align*}
This Hamiltonian represents a TQI when $V_z > \Delta$ and $\beta_1 < \beta_2$. From now on, we consider that $V_z = V_z (x)$ and $\Delta=\rm{const.}$ such that $V_z - \Delta= m(x)$ where $m(x)$ has a positive value inside the Majorana nanowires ($x \in (0,{\rm L})$) and $m(x)<0$ outside the nanowires. Furthermore, $\beta_1=\beta_2-\Delta\beta$ where $\Delta\beta(y) > 0$ once $y \in (0,\rm{W})$ and $\Delta\beta (y) <0$ outside the system given by dimensions $(\rm{L},\rm{W})$. Therefore, by choosing the signs of the parameters $V(x)$ (or equivalently $m(x)$) and $\Delta\beta(y)$, one can tune the system into a quadrupole phase. The Hamiltonian becomes
\begin{align*} 
\begin{split}
 H(\mathbf{k}) {}  = & V_z \tau_0 \sigma_z \eta_0 +  \Delta \tau_x \sigma_0 \eta_0 + \alpha k_x \tau_z \sigma_y \eta_0  -  \\
& \beta_2 k_y \tau_z \sigma_x \eta_x + \Delta\beta \tau_z \sigma_x \eta_y.
\end{split}
\end{align*}
By using this Hamiltonian, we can find the eigenstates localized on the edges (x or y) and use them to obtain edge Hamiltonians. For instance, the states localized on the x edge fulfill the ansatz $\Psi(x,k_y)= f(x) \psi_x(k_y)$ where $f(x)$ is a scalar and $\psi_x(k_y)$ is a spinor. The eigenvalue equation, after the Fourier transform (FT) \footnote{From now on, all Fourier transformed quantities are denoted with \fontfamily{ptm}\selectfont\texttildelow} in the x direction, is
\begin{align}\label{eq:Ham1} 
\begin{split}
 & (- i \alpha \frac{\partial_x f(x)}{f(x)} \tau_z \sigma_y \eta_0+  V_z \tau_0 \sigma_z \eta_0 +  \Delta \tau_x \sigma_0 \eta_0 ) \psi_x (k_y)   +  \\
& (-\beta_2 k_y \tau_z \sigma_x \eta_x + \Delta\beta \tau_z \sigma_x \eta_y) \psi_x (k_y) = \epsilon \psi_x(k_y).
\end{split}
\end{align}
The terms dependent on the x coordinate are grouped in the first parentheses and this equation only has a solution when all these terms give a constant. For simplicity we set this constant to be zero and obtain
\begin{equation} \label{eq:Hx1}
( - i \alpha \frac{\partial_x f(x)}{f(x)}  \tau_z \sigma_y \eta_0+  V_z \tau_0 \sigma_z \eta_0 +  \Delta \tau_x \sigma_0 \eta_0 )\psi_x (k_y) = 0.
\end{equation}
By denoting ${H}_x= - i \alpha \frac{\partial_x f(x)}{f(x)}  \tau_z \sigma_y \eta_0+  V_z \tau_0 \sigma_z \eta_0 +  \Delta \tau_x \sigma_0 \eta_0$, the equation $H_x \psi_x (k_y) = 0$ implies  $H_x^2 \psi_x (k_y) = 0$ or:
\begin{equation} \label{eq:Hx2}
[(- \alpha^2 (\frac{\partial_x f(x)}{f(x)})^2 +  V_z^2+  \Delta^2 )\mathbbm{1}_{8 \times 8}+ 2 V_z \Delta  \tau_x \sigma_z \eta_0 ] \psi_x (k_y) = 0.
\end{equation}
The unitary transformation that rotates $\tau_x \sigma_z \eta_0$ into a diagonal form is $U=\frac{1+i \tau_y}{\sqrt{2}} \sigma_0 \eta_0$. Thus, we get 
\begin{align} \label{eq:Hx3}
\begin{split}
 U^{\dagger}[ &  (- \alpha^2 (\frac{\partial_x f(x)}{f(x)})^2 +  V_z^2+ \\
&  \Delta^2 )\mathbbm{1}_{8 \times 8}+ 2 V_z \Delta  \tau_z \sigma_z \eta_0 ] U \psi_x (k_y) = 0.
\end{split}
\end{align}
The eigenvalue equation can easily be retrieved from the last expression. After imposing that the eigenvalues equal zero, we obtain $f(x)=\exp \left( \pm \int_0^x \frac{V_z(x') \pm \Delta}{\alpha} \right)$.
Now, let us concentrate on one of the solutions by inserting $f(x)= \exp \left( - \int_0^x \frac{V_z(x') - \Delta}{\alpha} \right)$ into Eq.~\eqref{eq:Hx1}. This choice of $f(x)$ corresponds to a state within the wire localized on the left boundary. The two corresponding eigenvectors are
\begin{align}\label{eq:X}
\begin{split}
& \psi_{x1} ^{\dagger}=
\frac{1}{2}  \left( {
\begin{array}{cccccccc}
-1 & 0 & 1 & 0 & 1 & 0 & 1 & 0
\end{array} } \right  ) \\
& \psi_{x2} ^{\dagger}=
\frac{1}{2}  \left( {
\begin{array}{cccccccc}
0 & -1 & 0 & 1 & 0 &1 & 0 & 1 
\end{array} } \right  ). 
\end{split}
\end{align}

The same procedure has to be performed  in the y direction to obtain the low-energy solutions. Using the ansatz $\Psi(k_x,y)=g(y) \psi_y (k_x)$ and replacing $k_y \rightarrow -i \partial_y$, we obtain the following eigenvalue equation:
\begin{align}\label{eq:Ham1b} 
\begin{split}
& (  i \beta_2 \frac{\partial_y g(y)}{g(y)} \tau_z \sigma_x \eta_x + \Delta\beta \tau_z \sigma_x \eta_y) \psi_y (k_x) + \\
 & (\alpha k_x \tau_z \sigma_y \eta_0+  V_z \tau_0 \sigma_z \eta_0 +  \Delta \tau_x \sigma_0 \eta_0 ) \psi_y (k_x)  = \epsilon \psi_y (k_x).
\end{split}
\end{align}
As before, the equation has a solution only if the terms in the first row give a constant and we choose it
to be zero. Then 
\begin{equation} \label{eq:Hy1}
(  i \beta_2 \frac{\partial_y g(y)}{g(y)}\tau_z \sigma_x \eta_x + \Delta\beta \tau_z \sigma_x \eta_y )\psi_y (k_x) = 0.
\end{equation}
Repeating the same procedure as in Eqs.~\eqref{eq:Hx1} and \eqref{eq:Hx2} and requiring eigenvalues of zero-energy, we find $g(y)= \exp \left(\pm \int_0^y dy' \frac{\Delta \beta (y')}{\beta_2} \right)$. By inserting $g(y)= \exp \left(\int_0^y dy' \frac{\Delta \beta (y')}{\beta_2} \right)$ in Eq.~\eqref{eq:Hy1}, we find eigenvectors that correspond to a state localized on the upper horizontal boundary of the system. They are solutions to a matrix equation: $(\mathbbm{1}_{8 \times 8}+ \mathbbm{1}_{2 \times 2} \otimes \mathbbm{1}_{2 \times 2} \otimes \eta_z)\psi_y =0  $:
\begin{align}\label{eq:Y}
\begin{split}
&\psi_{y1}^{\dagger}=
 \left( {
\begin{array}{cccccccc}
0 & 1 & 0 & 0 & 0 & 0 & 0 & 0 
\end{array} } \right  )  \\
&\psi_{y2}^{\dagger}=
 \left( {
\begin{array}{cccccccc}
0 & 0 & 0 & 1 & 0 & 0 & 0 & 0
\end{array} } \right  ) \\
& \psi_{y3}^{\dagger}=
 \left( {
\begin{array}{cccccccc}
0 & 0 & 0 & 0 & 0 & 1 & 0 & 0
\end{array} } \right  )  \\
& \psi_{y4}^{\dagger}=
 \left( {
\begin{array}{cccccccc}
0 & 0 & 0 & 0 & 0 & 0 & 0 & 1
\end{array} } \right  ). 
\end{split}
\end{align}

Finally, the edge Hamiltonians are obtained by projecting the remaining parts of Eqs.~\eqref{eq:Ham1} and \eqref{eq:Ham1b} with the projectors formed of states we have found. We are interested in their zero-energy states that have to be of the same form
if there exists a simultaneous eigenstate of both domain walls.

The low-energy Hamiltonian on the x edge is extracted from Eq.~\eqref{eq:Ham1}, using the subspace spanned by the eigenvectors $(\psi_{x1},\psi_{x2})$ to form a projector $P_x=\ket{\psi_{x1}} \bra{\psi_{x1}} + \ket{\psi_{x2}} \bra{\psi_{x2}}$. The projected edge Hamiltonian reads
\begin{equation}
H_{\rm{edge},x} = \beta_2 k_y \rho_x - \Delta \beta \rho_y,
\end{equation}
where $\rho_i$ are Pauli matrices in the basis $(\psi_{x1},\psi_{x2})$.

After the FT, the resulting eigenvalue equation is ${H}_{\rm{edge},x} \; \phi_y = 0$. Here, $\phi_y= g'(y)\psi_{yx}$ where $g'(y)$ is a scalar function which has all the spatial dependence and $\psi_{yx}$ is a spinor.
The spectrum of ${H}_{\rm{edge},x}$ is ${E}^2_{\rm{edge},x} =-\beta_2^2 (\frac{\partial_y g'(y)}{g'(y)})^2 + (\Delta \beta)^2$ and from setting it to zero, we determine $g'(y)$ as $g'(y)=g(y)= \exp \left(\pm \int_0^y dy' \frac{\Delta \beta (y')}{\beta_2} \right)$.
Again, by choosing the solution that describes the state localized on the upper boundary of the edge and inserting it into the eigenvalue equation, one gets the matrix equation
\begin{equation}
(\mathbbm{1}_{2 \times 2}+\rho_z) \psi_{yx} = 0,
\end{equation}
and the eigenvector corresponding to a negative eigenvalue is $\psi_{xy}^{\dagger} = (0 \; \; 1)$. This indicates that in the basis $(\psi_{x1},\psi_{x2})$, the spinor of the corner mode is of the form $\psi_{x2}$ or equivalently
\begin{equation}\label{eq:corner}
\psi_{\rm{corner}}^{\dagger} =  
\frac{1}{2}  \left( {
\begin{array}{cccccccc}
0 & -1 & 0 & 1 & 0 &1 & 0 & 1 
\end{array} } \right  ).
\end{equation}

For the other edge, the remaining part of Eq.~\eqref{eq:Ham1b} is projected using the states defined in Eq.~\eqref{eq:Y}, by forming the operator $P_y=\ket{\psi_{y1}} \bra{\psi_{y1}} + \ket{\psi_{y2}} \bra{\psi_{y2}}+ \ket{\psi_{y3}} \bra{\psi_{y3}}+ \ket{\psi_{y4}} \bra{\psi_{y4}}$.
The resulting Hamiltonian is
\begin{equation} \label{eq:Py}
 H_{\rm{edge},y}=
 \left( {
\begin{array}{cccc}
V_z & - i \alpha k_x & \Delta & 0  \\
i \alpha k_x & -V_z & 0 & \Delta  \\
\Delta & 0 & V_z & i \alpha k_x \\
0 & \Delta & -i\alpha k_x & -V_z 
\end{array} } \right  ), 
\end{equation}
that is equivalent to
\begin{equation} 
H_{\rm{edge},y} = \alpha k_x \theta_z \gamma_y+ V_z \gamma_z + \Delta \theta_x, 
\end{equation}
where the Pauli matrices $\gamma_i$ and $\theta_i$ act in the space spanned by $(\psi_{y1},\psi_{y2},\psi_{y3},\psi_{y4})$.
After the FT, the eigenvalue equation becomes: 
\begin{equation} \label{eq:HedgeY}
(-i \alpha \partial_x \theta_z \gamma_y+ V_z \gamma_z + \Delta \theta_x) \phi_x= 0,
\end{equation}
where $\phi_x = f'(x) \psi_{xy}$.
Then:
\begin{equation}
{H}^2_{\rm{edge},y} = (- \alpha^2 (\frac{\partial_x f'(x)}{f'(x)})^2 +  V_z^2+  \Delta^2 )\mathbbm{1}_{4 \times 4}+ 2 V_z \Delta  \theta_x \gamma_z.
\end{equation}

In order to obtain a diagonal matrix, we rotate an off-diagonal term with a unitary transformation $U=\frac{1+i \theta_y}{\sqrt{2}} \gamma_0$. The resulting eigenvalues are ${E}^2_{\rm{edge},y} = - \alpha^2 (\frac{\partial_x f'(x)}{f'(x)})^2 +  V_z^2+  \Delta^2  \pm 2 V_z \Delta$. As before, $f'(x)$ is determined from the requirement that these eigenvalues are equal to zero and $f'(x)=f(x)= \exp \left( \pm \int_0^x \frac{V_z(x') \pm \Delta}{\alpha} \right)$. By choosing $f(x)$ corresponding to the state localized on the left boundary and inserting it into the eigenvalue equation Eq.~\eqref{eq:HedgeY}, we get the corresponding spinor $\psi^{\dagger}_{xy}=(-1 \; \; 1 \; \; 1 \; \; 1)$. In the basis  $(\psi_{y1},\psi_{y2},\psi_{y3},\psi_{y4})$, this implies:
\begin{equation}\label{corner2}
\psi_{\rm{corner}}^{\dagger} =  -\psi_{y1}^{\dagger} + \psi_{y2}^{\dagger} + \psi_{y3}^{\dagger} + \psi_{y4}^{\dagger}, 
\end{equation}
that has the same form, up to a normalization constant, as the spinor in Eq.~\eqref{eq:corner}.
Therefore, we have proven that there exists a corner state shared by the two edge Hamiltonians:
\begin{equation*} 
\begin{split}
&\Psi^{\rm{corner}} (x,y) = \\
& \exp \left(- \int_0^x dx' \frac{V_z(x')-\Delta}{\alpha} \right) \exp \left( \int_0^y dy' \frac{\Delta \beta (y')}{\beta_2} \right) \psi_{\rm{corner}},
\end{split}
\end{equation*}
where 
\begin{equation*}
\psi_{\rm{corner}}^{\dagger}= \frac{1}{2}
 \left( {
\begin{array}{cccccccc}
0 & -1 & 0 & 1 & 0 & 1 & 0 & 1
\end{array} } \right).
\end{equation*}

\bibliography{AHOTI}
\clearpage